\documentclass[]{aa}  

\bibpunct{(}{)}{;}{a}{}{,} 
\usepackage{graphicx}
\usepackage[varg]{txfonts}
\usepackage{comment}
\usepackage{gensymb}

\usepackage{hyperref}
\hypersetup{%
  colorlinks=true,
  linkcolor=blue,
  urlcolor=blue,
  citecolor=blue,
  bookmarksnumbered=true,
  bookmarksopenlevel=1,
  }

\begin{document} 

\title{Oscillatory reconnection and resonant response to wave excitation in 2D coronal null points}

\author{%
{Konstantinos Karampelas}\inst{\ref{aff:CmPA},\ref{aff:CFSA}} \orcid{0000-0001-5507-1891}
\and {Tom Van Doorsselaere}\inst{\ref{aff:CmPA}} \orcid{0000-0001-9628-4113}
}

\institute{%
\label{aff:CmPA}{Centre for mathematical Plasma Astrophysics, Department of Mathematics, KU Leuven, Celestijnenlaan 200B, 3001 Leuven, Belgium.}
\and
\label{aff:CFSA}{Centre for Fusion, Space and Astrophysics, Department of Physics, University of Warwick, Coventry CV 4 7AL, UK.}
\email{konstantinos.karampelas@warwick.ac.uk}\\
}

\date{Received \today; Accepted \today}
 
\abstract
{Null points are magnetic field singularities, where the magnetic field strength rapidly drops to zero. In the solar atmosphere, null points are known sites of magnetic reconnection and wave generation and are associated with highly energetic phenomena, such as flares.}
{The aim of this study is to explore the connection between the properties of oscillatory reconnection at null points and the latter's nature as resonant cavities for waves.}
{We perform a set of 2D and 2.5D magnetohydrodynamics simulations of single null points in a stratified solar atmosphere, using the PLUTO code.}
{We perturb each null point through a single propagating pulse and its reflections from the bottom boundary, hitting the null point in an asymmetrical fashion. This leads to both periodic reconnection events and wave refraction around the null point. We find that each null point imposes frequencies on the reconnection matching those of the waves generated from the individual resonant cavity. These frequencies also differ from those excited by the low frequency driver of the reflected waves returning to the null point, the latter lying outside the $95\%$ confidence interval. As such, excited periodic reconnection can be characterised as oscillatory reconnection, i.e. with properties intrinsic to the null points. Finally, the generated waves at the null propagate across the domain, reminiscent of the observed quasi-periodic fast-propagating waves.}
{We provide results showing a direct connection between oscillatory reconnection and the generated high-frequency wavetrains at null points in the solar corona. The propagating waves generated at the resonant cavity, reminiscent of the observed quasi-periodic fast-propagating waves can provide us a diagnostic tool for the reconnection process at the null point and the coronal plasma conditions.}


\titlerunning{Oscillatory reconnection and resonance at coronal null points}
\authorrunning{Karampelas \& Van Doorsselaere}

\maketitle
\nolinenumbers
\section{Introduction} \label{sec:introduction}

Null points are singularities of the magnetic field, i.e. areas where the field amplitude goes to zero. We know from potential field extrapolations of photospheric magnetograms that null points are abundant in the solar atmosphere \citep[e.g.][]{BrownPriest2001AnA,Longcope2005LRSP,Regnier2008AnA}. Numerous studies in the past have dealt with the interaction of null points and the ubiquitous waves in the solar atmosphere \citep[see][for a review]{McLaughlin2011SSRv}, such as wave refraction and mode conversion \citep[e.g.][]{McLaughlin2004,McLaughlin2006b,Gruszecki2011null,Thurgood2012, Santamaria2015}. A recent study \citep{KumarPankaj2024NatCo..15.2667K} showed the direct extreme ultraviolet (EUV) imaging of mode conversion from a fast-mode to a slow-mode MHD wave near a 3D null point using observations from the Solar Dynamics Observatory/Atmospheric Imaging Assembly (SDO/AIA). Null points are also known to behave as resonant cavities \citep{Santamaria2018}, generating waves at frequencies imposed by the background plasma conditions. An example of this are the high-frequency wave trains ($\sim 80$\,mHz) reported in numerical studies by \citet{Santamaria2016} and \citet{Santamaria2017}. Null points were also shown to act as sources of coronal jets, slow, fast and Alfv\'en waves  \citep[e.g.][]{2014SoPh..289.3043L, 2017ApJ...834...62K, Thurgood2017ApJ, 2018ApJ...862....6C} when subjected to external perturbations and driving, stressing the importance of wave$-$null point interactions for coronal seismology.

\begin{figure*} [t]
    \centering
    \resizebox{\hsize}{!}{
    \includegraphics[trim={3.cm 0.3cm 3.8cm 1.4cm},clip,scale=0.55]{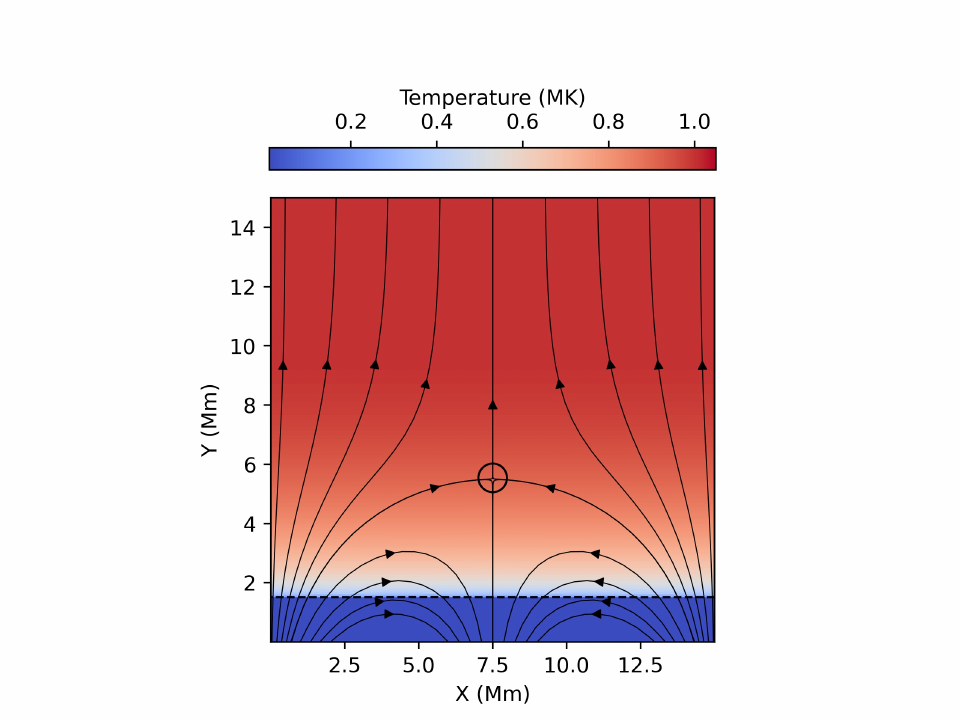}
    \includegraphics[trim={3.9cm 0.3cm 3.8cm 1.4cm},clip,scale=0.55]{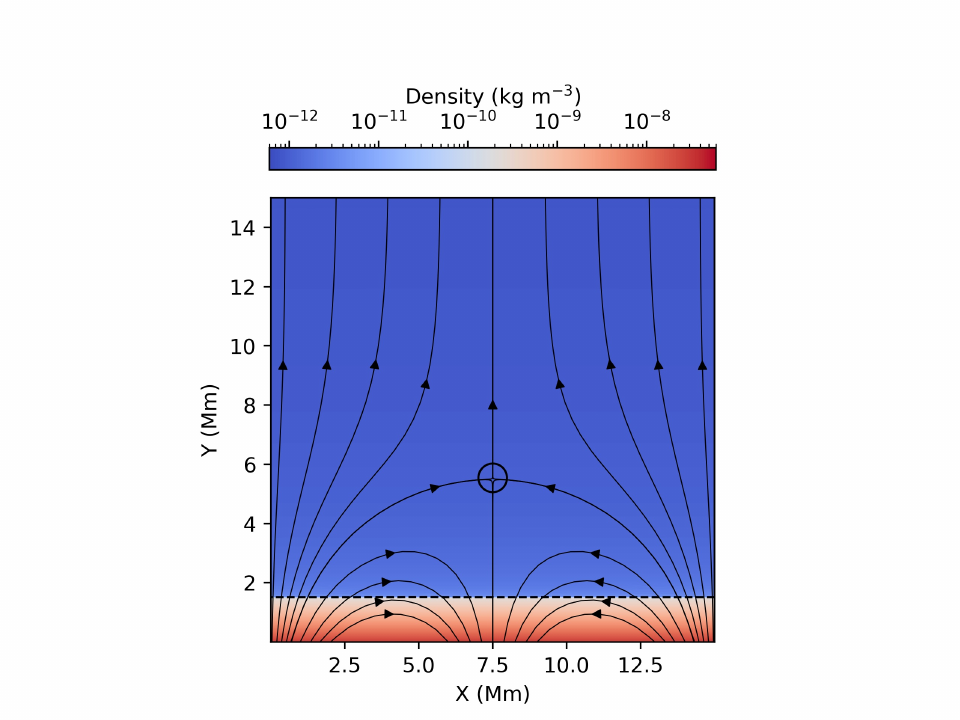}
    \includegraphics[trim={3.9cm 0.3cm 3.8cm 1.4cm},clip,scale=0.55]{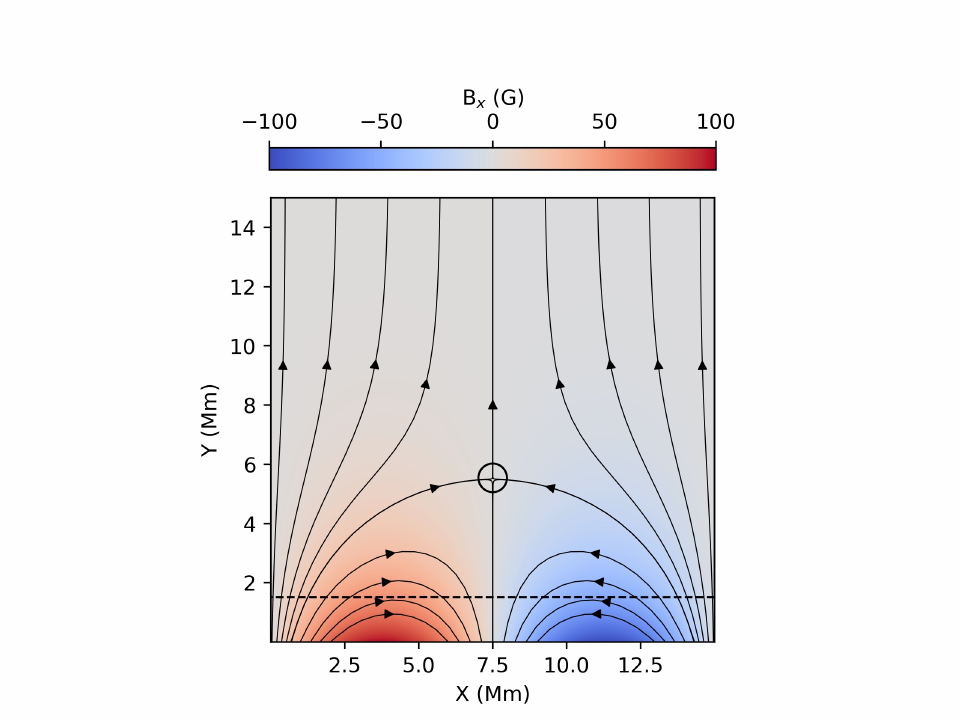}
    \includegraphics[trim={3.9cm 0.3cm 3.8cm 1.4cm},clip,scale=0.55]{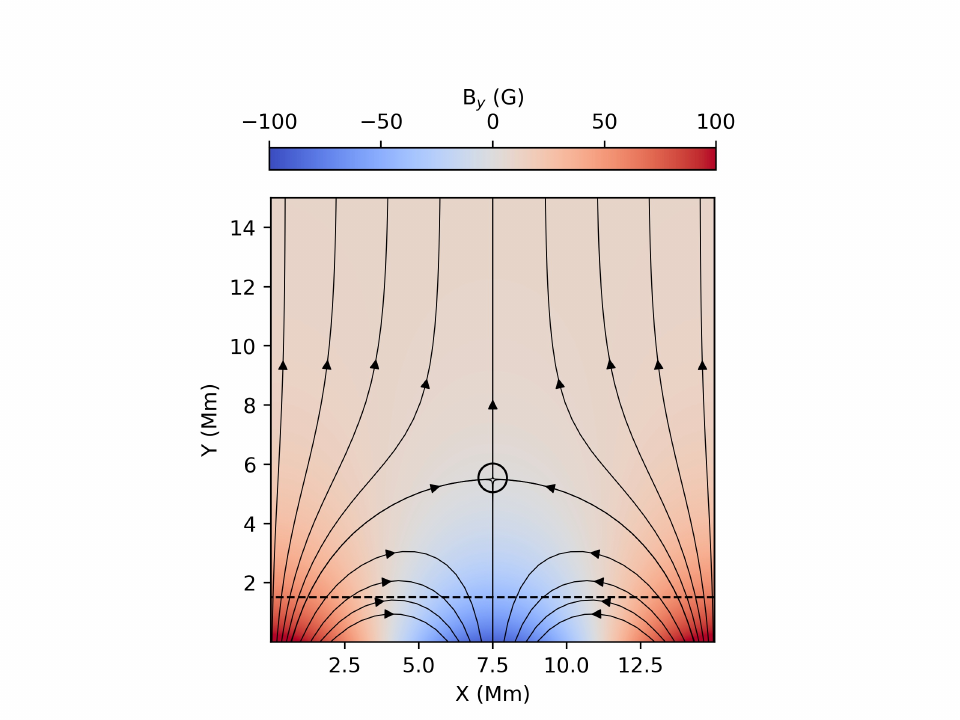}
    }
    \caption{Profiles of the temperature, density and planar magnetic field components for the default null point model (M1), over-plotted with the magnetic field lines. The dashed black line is the top of the transition region (where $T=0.1$ MK). The black circle is the $\beta =1$ layer.} \label{fig:inicon}
\end{figure*}

Magnetic null points are also a preferred locations for magnetic reconnection \citep{Parker1957JGR,Sweet1958IAUS,Petschek1964NASSP} in the solar atmosphere \citep[e.g.][]{PriestForbes2000book, Heggland2009ApJ, 2011AdSpR..47.1508P, PontinPriest2022LRSP...19....1P,2023NatCo..14.2107C}, leading to plasma heating and wave generation \citep[e.g.][]{2006A&A...452..343N, 2008ApJ...683L..83N, LongcopePriest2007PhPl...14l2905L, 2009ApJ...705L.217H, Thurgood2017ApJ, Mondal2024ApJ...977..235M, Srivastava2025ApJ...984...36S}. Reconnection can cause shock heating, mass ejection, and particle acceleration, making it the central engine behind highly energetic phenomena such as solar flares \citep[e.g.][]{ShibataMagara2011LRSP,2015ApJ...812..105J} and therefore, making null points preferential locations for flares to occur \citep[e.g.][]{2011A&A...533A..18M}. In this context, periodic reconnection at magnetic null points has been proposed as an explanation for the observed quasi-periodic pulsations (QPPs) of solar \citep[e.g.][]{Kupriyanova2016SoPh,VanDoorsselaere2016SoPh,2020ApJ...893....7L,2021ApJ...921..179L, 2022FrASS...932099L, 2021ApJ...910..123C, 2022ApJ...931L..28L, 2022RAA....22j5017S} and stellar flares \citep[e.g.][]{2019A&A...629A.147B,2019ApJ...884..160V,2020A&A...636A..96M, 2021SoPh..296..162R}. In addition, periodically driven reconnection in null points has also been shown to drive spicule like jets in simulations of the lower solar atmosphere, with lengths and lifetimes that match observations \citep{Heggland2009ApJ}.

A form of periodic reconnection that has been included in reviews summarizing our current understanding of QPPs \citep{McLaughlin2018SSRv, Kupriyanova2020STP, Zimovets2021SSRv} is the mechanism of oscillatory reconnection \citep{CraigMcClymont1991ApJ}. This mechanism is characterised by periodic changes in the magnetic connectivity and topology of the field, where the periodicity is imposed by the properties of the null-point itself, rather than that of an external driver. The mechanism has been studied numerically in 2D and 3D, in isolated null points \citep{McLaughlin2009,McLaughlin2012A&A,Thurgood2017ApJ} as well as in a stratified solar atmosphere \citep{2009A&A...494..329M,McLaughlin2012ApJ, 2025arXiv250524335W}. 

Another phenomenon that has been associated with periodic and/or oscillatory magnetic reconnection, is that of the quasi-periodic fast-propagating (QFP) magnetosonic waves. In \citet{LiLepingQFPW2018ApJ...868L..33L}, QFP waves were detected in a region of coronal condensations caused by magnetic reconnection between open and closed coronal loops. The authors of that study suggested that periodic or oscillatory reconnection was responsible for these waves. Oscillatory reconnection has also been considered as a possible driver of quasi-periodic fast-propagating (QFP) magnetosonic waves from an erupting flux rope, such as the one observed in \citet{2018ApJ...853....1S}. 

The connection between QFP waves and QPPs, and therefore their mutual connection to periodic reconnection in flares has also been suggested in past studies, highlighting the importance of further exploring this possible mutual connection. In \citet{LiuQFPW2011ApJ...736L..13L}, QFP wave trains were detected in EUV by SDO/AIA in a flare and coronal mass ejection event, suggesting a common origin between the two phenomena. In addition, \citet{KumarQFPW2017ApJ...844..149K} have reported quasi-periodic bursts in radio, microwave and soft X-ray emission associated with observed QFP waves of similar instant periods, and both were theorised to be generated from quasi-periodic magnetic reconnection in the cusp region above flaring loops.

The seismological implications of the periodicity of oscillatory reconnection have been investigated in a series of recent studies focusing on the effects of thermal conduction \citep{Karampelas2022a}, the resistivity \citep{Talbot2024ApJ...965..133T} and the strength and form of external excitation on the periodicity of the mechanism \citep{Karampelas2022b}. In \citet{Karampelas2023ApJ...943..131K}, a semi-empirical formula was derived, describing the relation of the reconnection period to the magnetic field, density and temperature of the background plasma, while self-similarity in the solutions of oscillatory reconnection for the temperature of the generated reconnection jets, current density and Ohmic heating was reported in \citet{Schiavo2024ApJ...975...10S}.

While wave generation was reported in \citet{Santamaria2017} and \citet{Santamaria2018} for a resonant cavity that included a null point in the solar atmosphere, the reconnection process was not considered in those studies. Similarly, following the results of \citet{KumarPankaj2024NatCo..15.2667K}, \citet{ZhongYu2026ApJ..1004...28Z} studied the efficiency of fast-to-slow mode conversion in a full 2D MHD numerical model of a null point in a pseudostreamer. That study showed a power-law dependence with an index of approximately $1.06$ for the mode conversion efficiency with respect to the width of the initial velocity perturbation, while only a weak dependence on the initial amplitude was derived. However, that study did not explore magnetic reconnection at the null point, or the relation between the periodicity of the reconnection and the generated waves from the resonant cavity. On the other hand, wave generation from reconnection at a null point has been reported in the past \citep[e.g.][]{Thurgood2017ApJ, Mondal2024ApJ...977..235M}, in uniform coronal plasma without gravity. In a recent study of an isolated 3D null point in uniform plasma without gravity \citep{Schiavo2026ApJ...999...50S}, it was shown that oscillatory reconnection generates a slow magnetoacoustic wave of the same period that propagates outward along the spine and fan plane, as well as an Alfvén wave of the same period that period that propagates exclusively along the y-axis in the fan plane. Additional studies of a coronal null point \citep[e.g.][]{Tarr2017ApJ...837...94T, Tarr2019ApJ...879..127T} have focused on periodic reconnection driven by external waves, but no oscillatory reconnection (i.e. with a self-imposed periodicity) nor a resonant cavity have been observed in those studies. In this current paper, we explore the connection between the periodicity of oscillatory reconnection and the wave generation properties of the resonant cavity associated with a coronal null point, as well as the relation of both to outward propagating QFPs. In section \ref{sec:setup} we introduce the numerical model of a null point embedded in a stratified solar atmosphere and the different 2D and 2.5D setups that we are using in our study. Section \ref{sec:results} contains the results of our numerical simulations, while the overall discussion and our conclusions are presented in Section \ref{sec:discussions}.

\section{Numerical setup} \label{sec:setup}

\begin{figure*}
    \centering
    \resizebox{\hsize}{!}{
    \includegraphics[trim={0.cm 1.2cm 4.cm 0.cm},clip,scale=0.55]{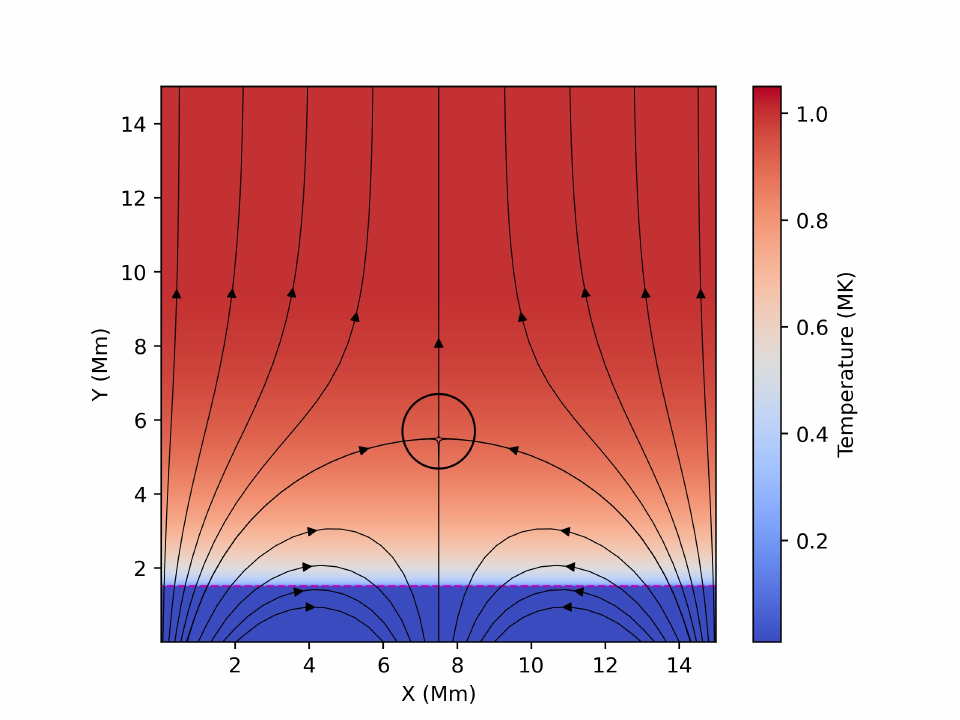}
    \includegraphics[trim={2.5cm 1.2cm 4.cm 0.cm},clip,scale=0.55]{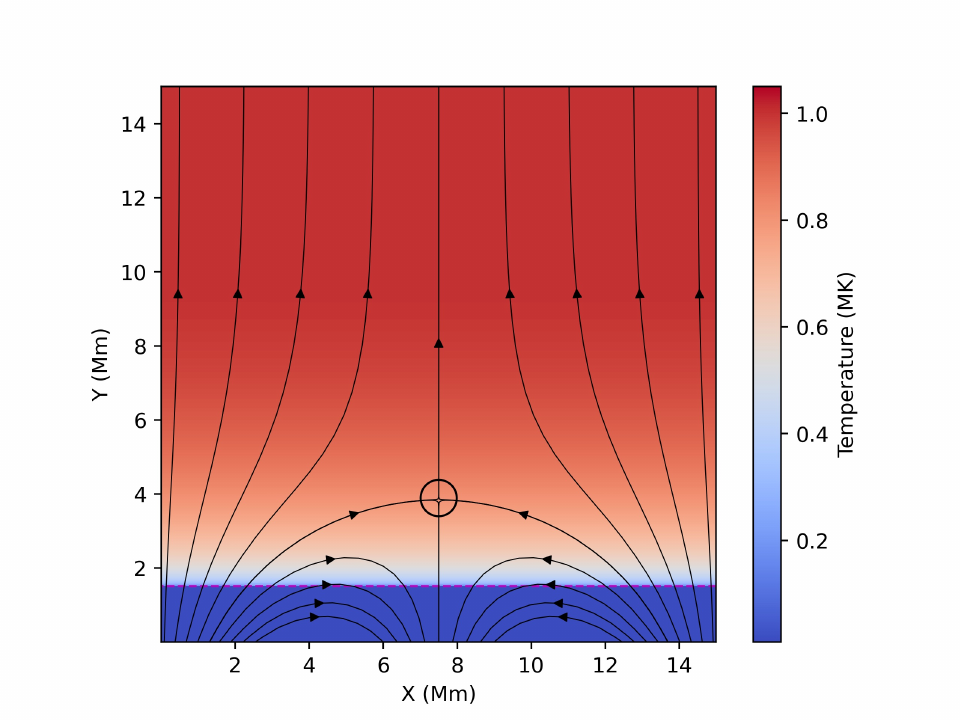}
    \includegraphics[trim={2.5cm 1.2cm 4.cm 0.cm},clip,scale=0.55]{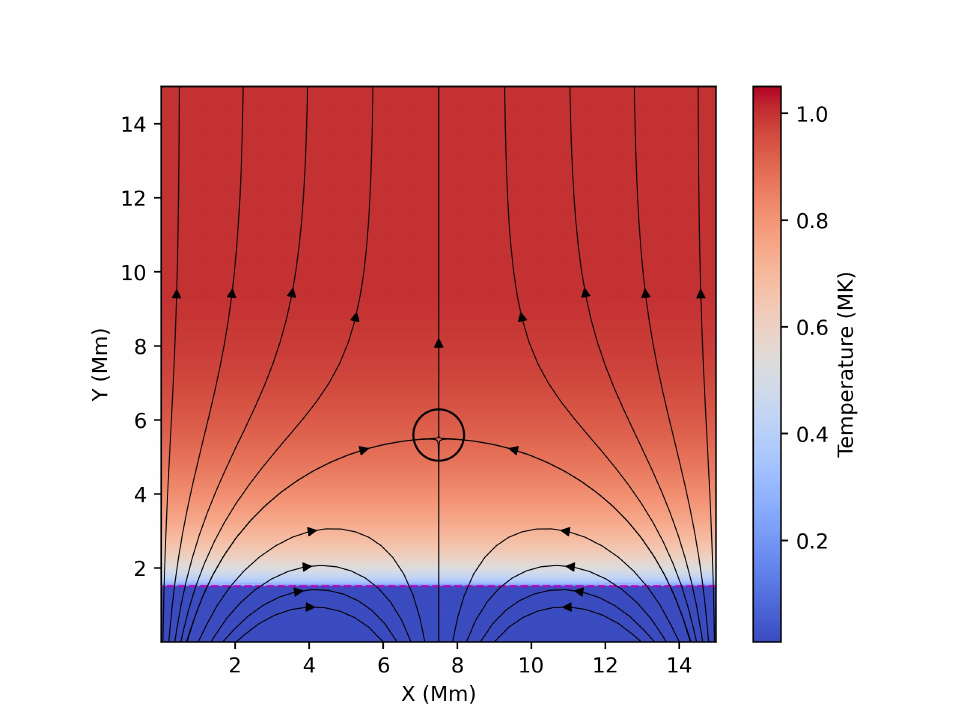}
    \includegraphics[trim={2.5cm 1.2cm 0.cm 0.cm},clip,scale=0.55]{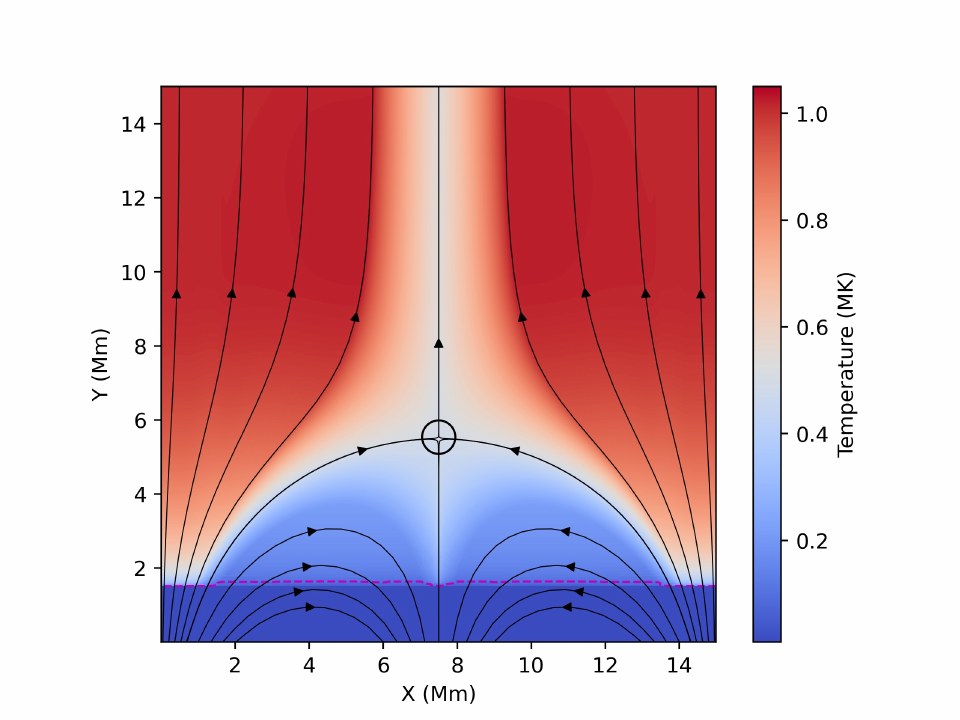}}
    
    \resizebox{\hsize}{!}{
    \includegraphics[trim={0.cm 0.cm 4.cm 1.4cm},clip,scale=0.55]{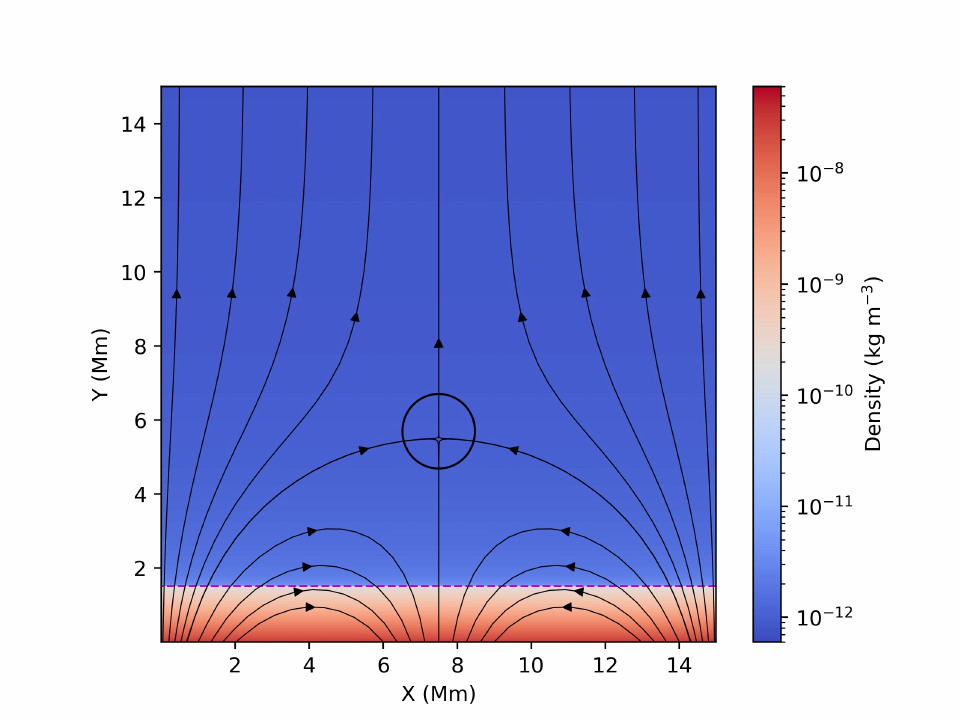}
    \includegraphics[trim={2.5cm 0.cm 4.cm 1.4cm},clip,scale=0.55]{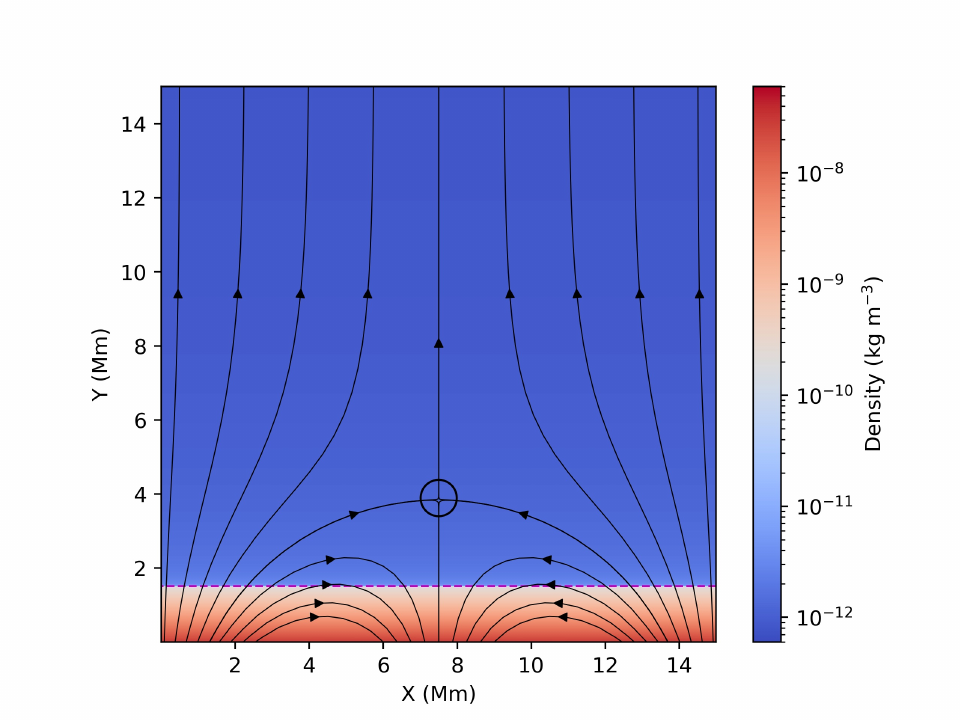}
    \includegraphics[trim={2.5cm 0.cm 4.cm 1.4cm},clip,scale=0.55]{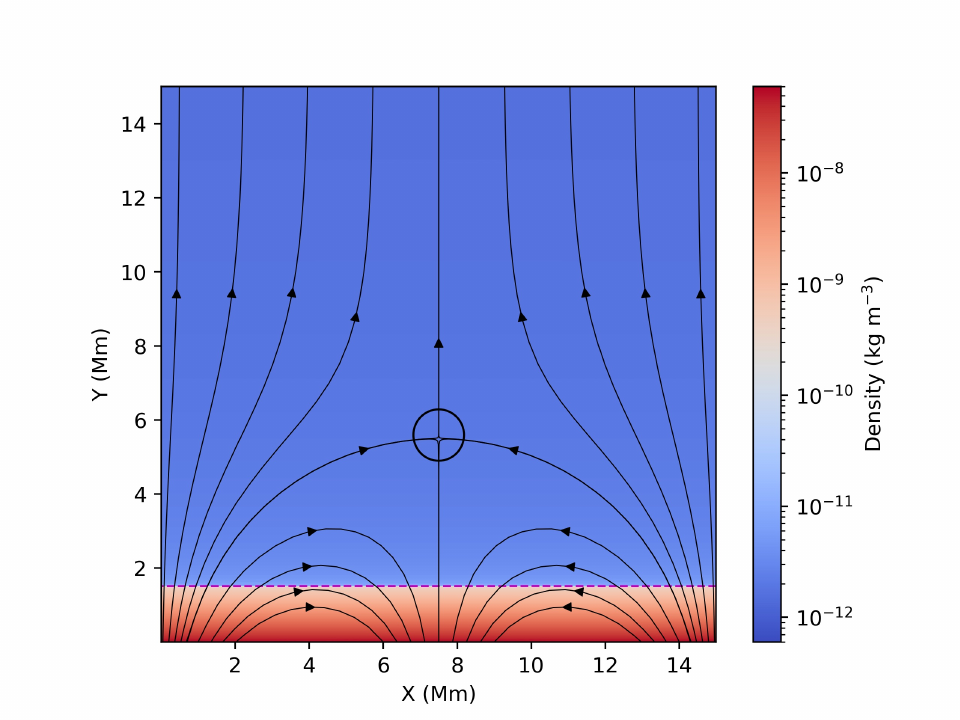}
    \includegraphics[trim={2.5cm 0.cm 0.cm 1.4cm},clip,scale=0.55]{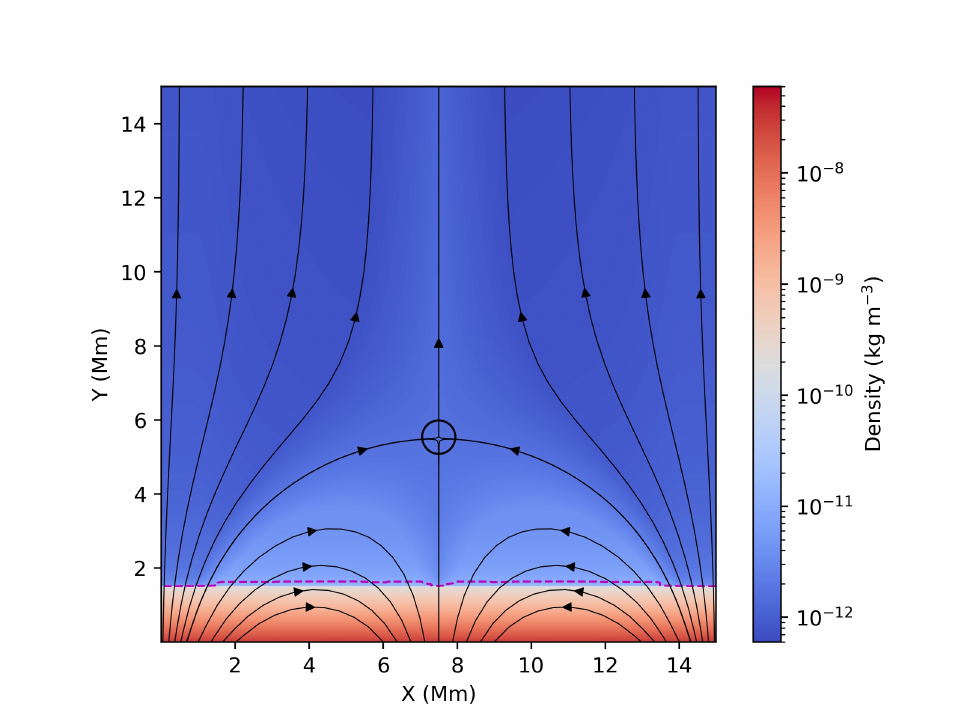}}
    \caption{Profiles of the temperature (top row) and density (bottom row), over-plotted with the magnetic field lines, for models M2, M3, M5 and M6, starting from the left. The transition region (where $T=0.1$ MK) is plotted as a dashed purple line.}    \label{fig:alternative}
\end{figure*}

\begin{figure*}
    \centering
    \resizebox{\hsize}{!}{
    \includegraphics[trim={0.cm 1.2cm 4.cm 0.cm},clip,scale=0.55]{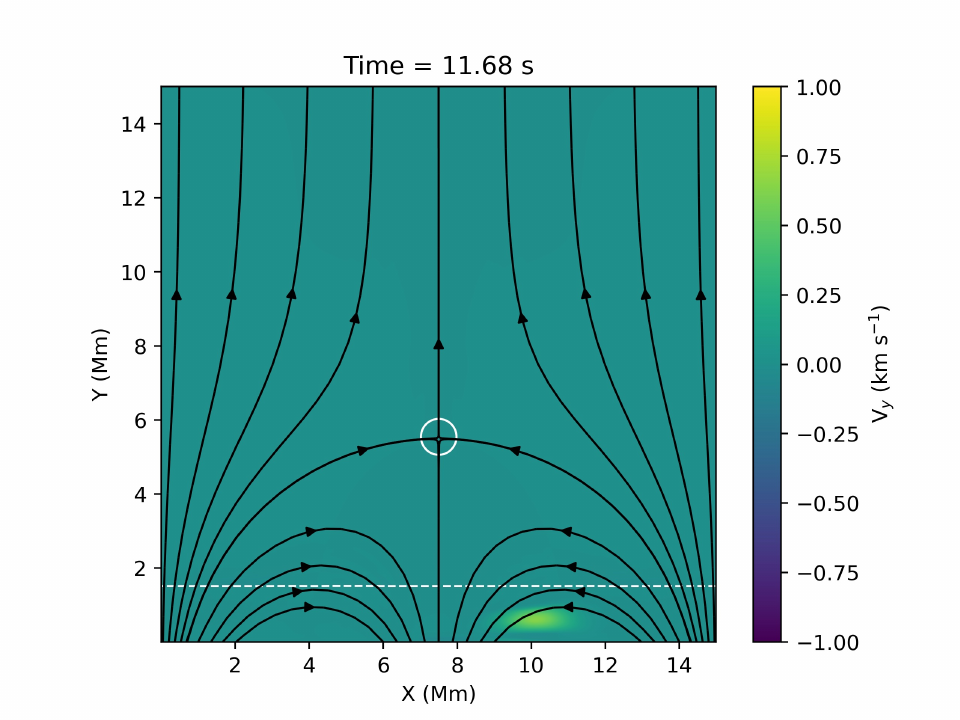}
    \includegraphics[trim={2.5cm 1.2cm 4.cm 0.cm},clip,scale=0.55]{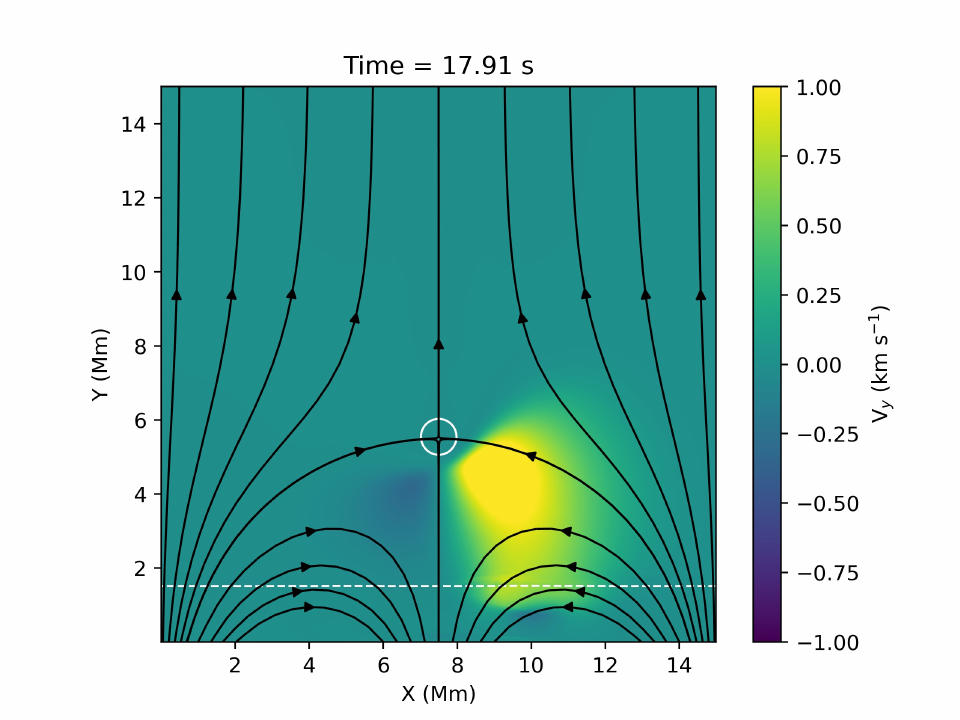}
    \includegraphics[trim={2.5cm 1.2cm 4.cm 0.cm},clip,scale=0.55]{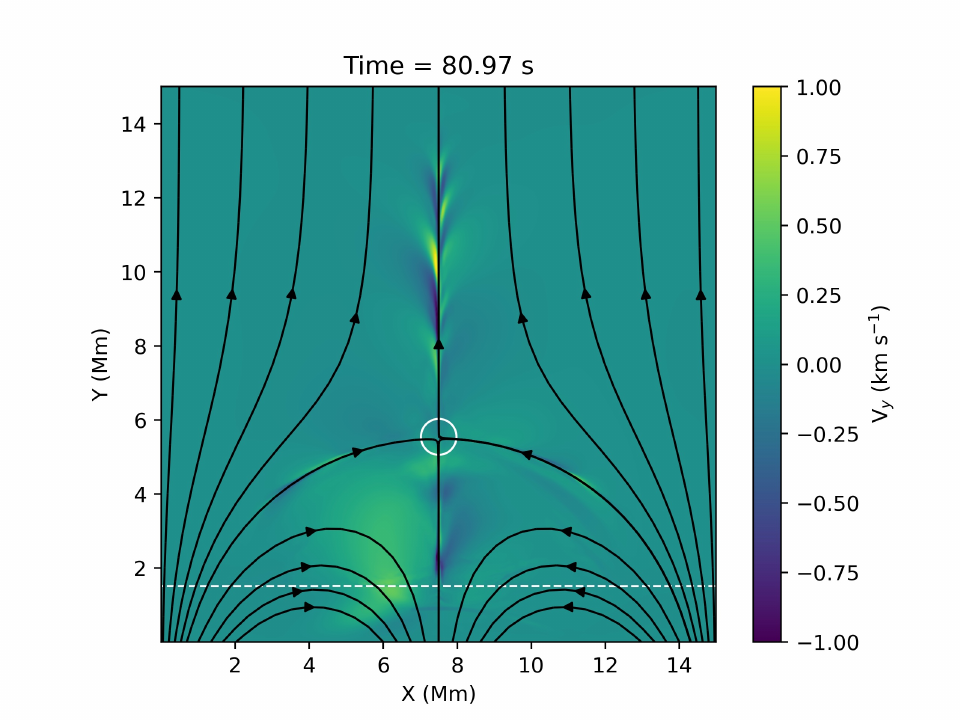}
    \includegraphics[trim={2.5cm 1.2cm 0.cm 0.cm},clip,scale=0.55]{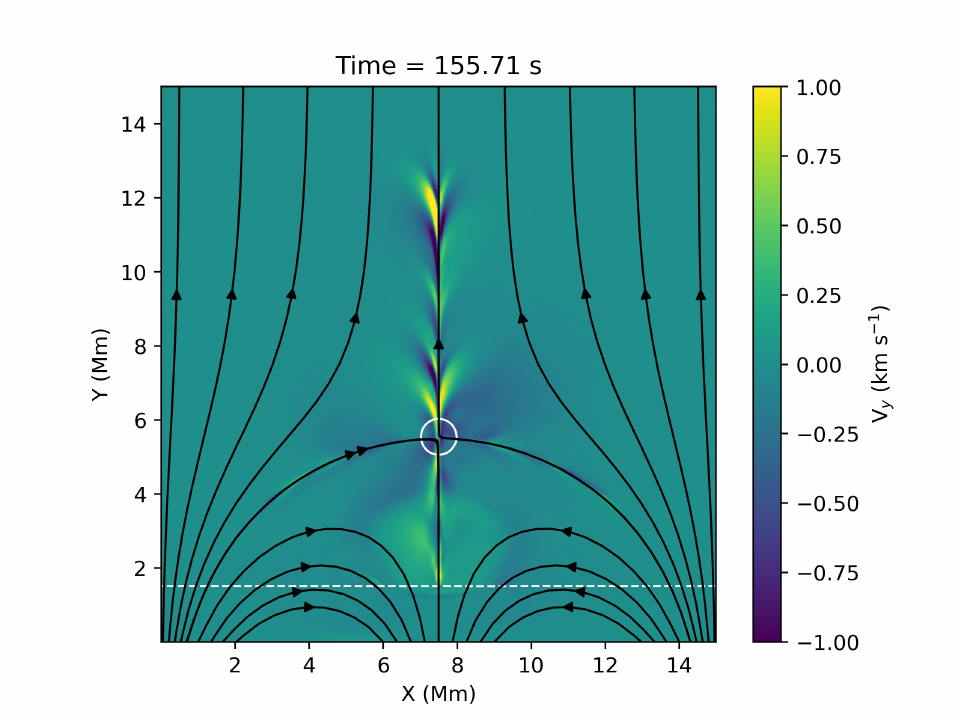}
    }
    \resizebox{\hsize}{!}{
    \includegraphics[trim={0.cm 0.cm 4.cm 1.3cm},clip,scale=0.55]{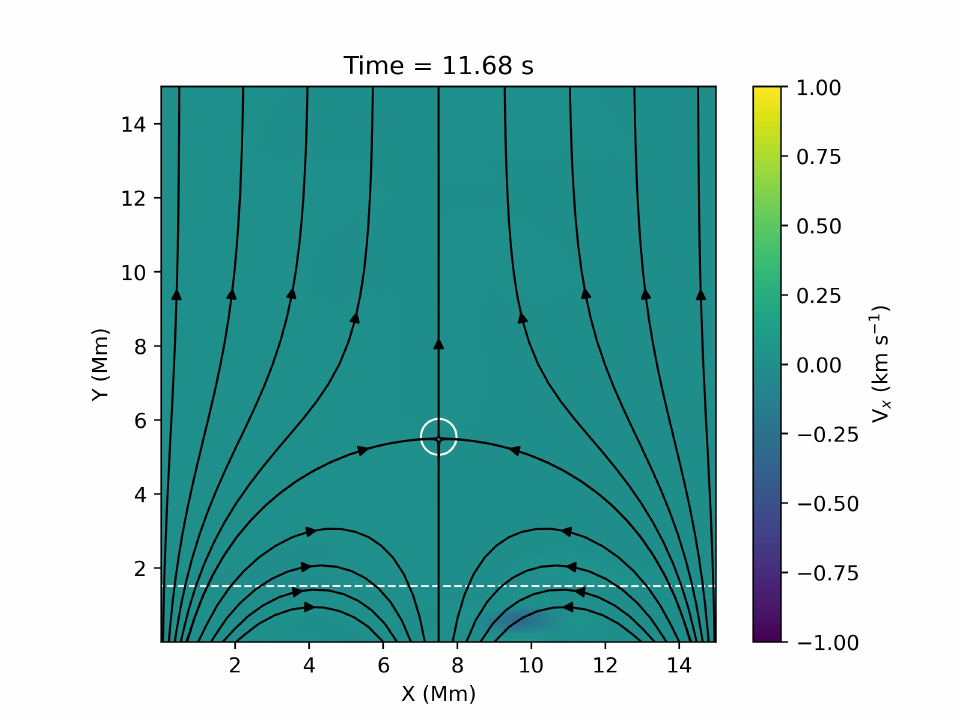}
    \includegraphics[trim={2.5cm 0.cm 4.cm 1.3cm},clip,scale=0.55]{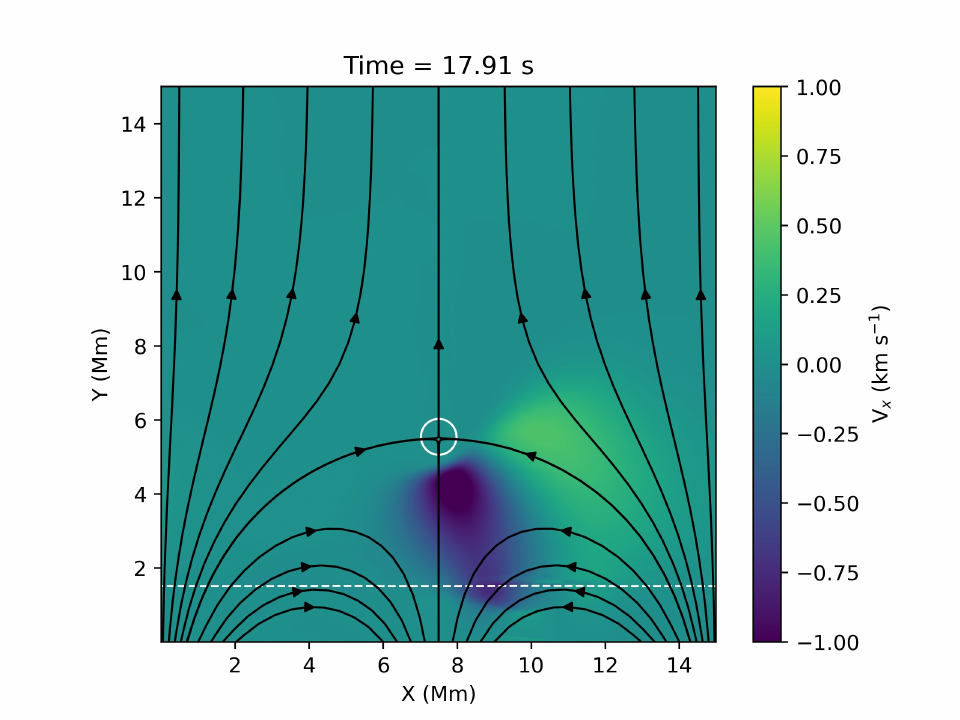}
    \includegraphics[trim={2.5cm 0.cm 4.cm 1.3cm},clip,scale=0.55]{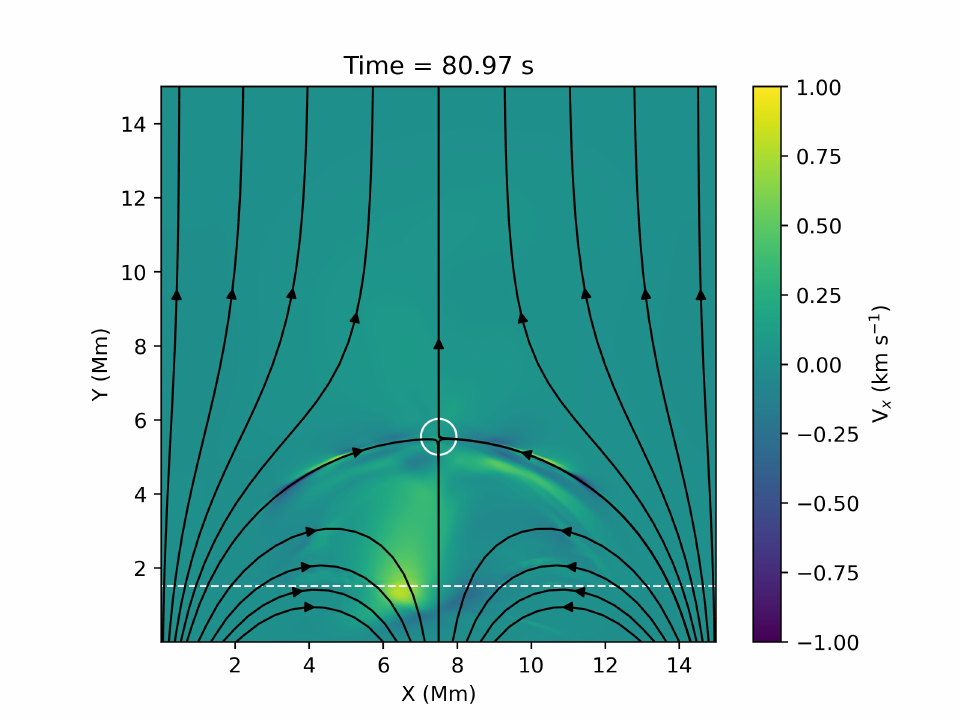}
    \includegraphics[trim={2.5cm 0.cm 0.cm 1.3cm},clip,scale=0.55]{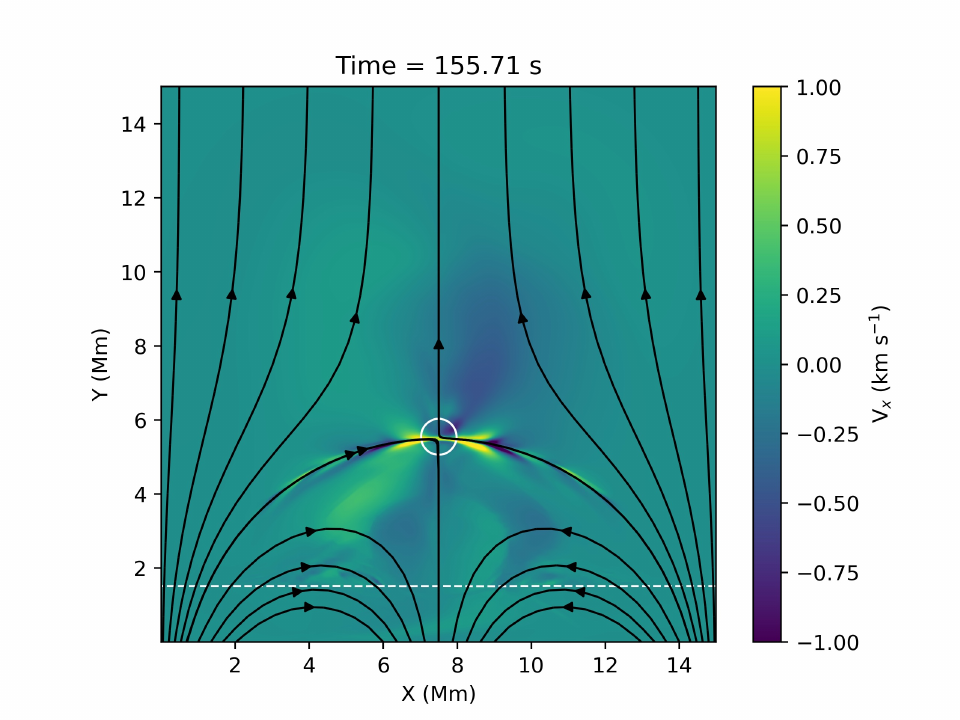}
    }
    \caption{Snapshots of the $V_y$ (top row) and $V_x$ (bottom row) velocity components, for M1. The $\beta=1$ layer (white circle), top of the transition region (white dashed line) and magnetic field lines (black arrows) are also shown. Accompanying animations for the two velocities can be found in the online version of this manuscript.} \label{fig:2Dvelocities}
\end{figure*}

\begin{figure*}
    \centering
    \resizebox{\hsize}{!}{
    \includegraphics[trim={0.cm 0.9cm 1.cm 0.cm},clip,scale=0.55]{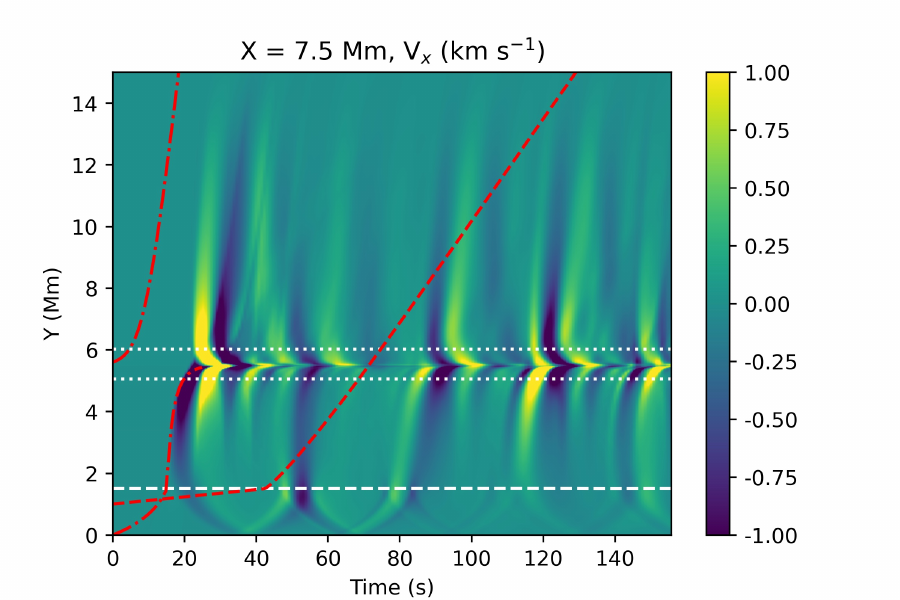}
    \includegraphics[trim={1.cm 0.9cm 3.5cm 0.cm},clip,scale=0.55]{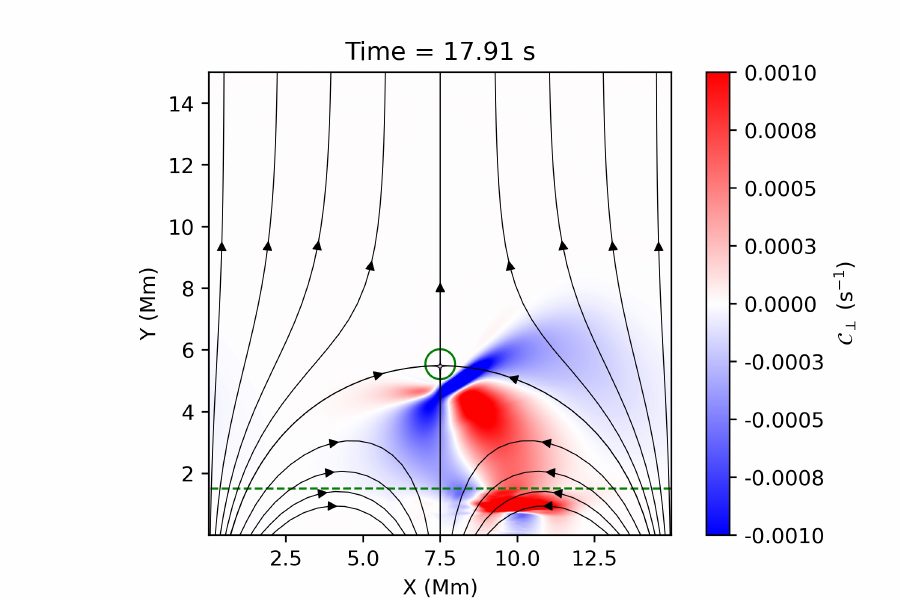}
    \includegraphics[trim={3.5cm 0.9cm 0.cm 0.cm},clip,scale=0.55]{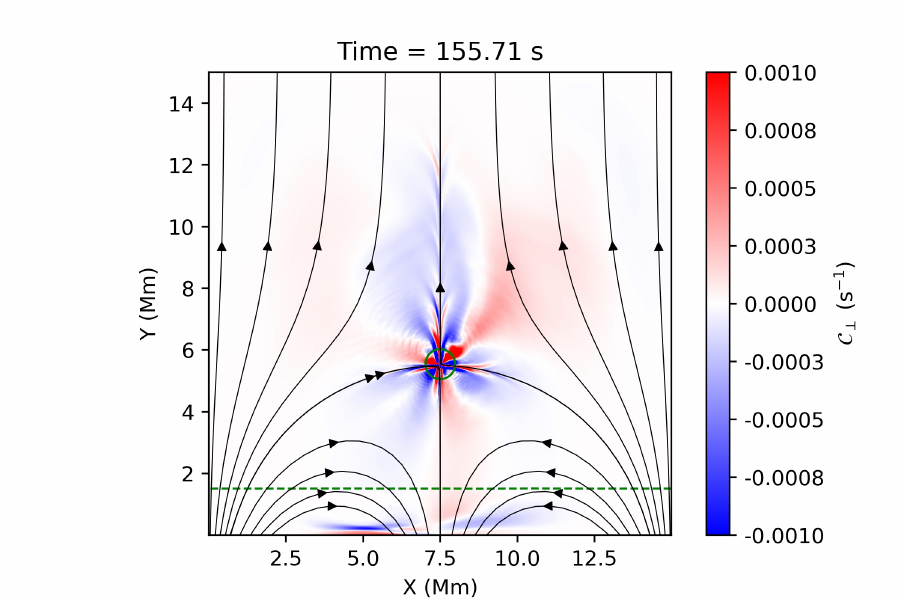}
    }
    \resizebox{\hsize}{!}{
    \includegraphics[trim={0.cm 0.cm 1.cm 0.5cm},clip,scale=0.55]{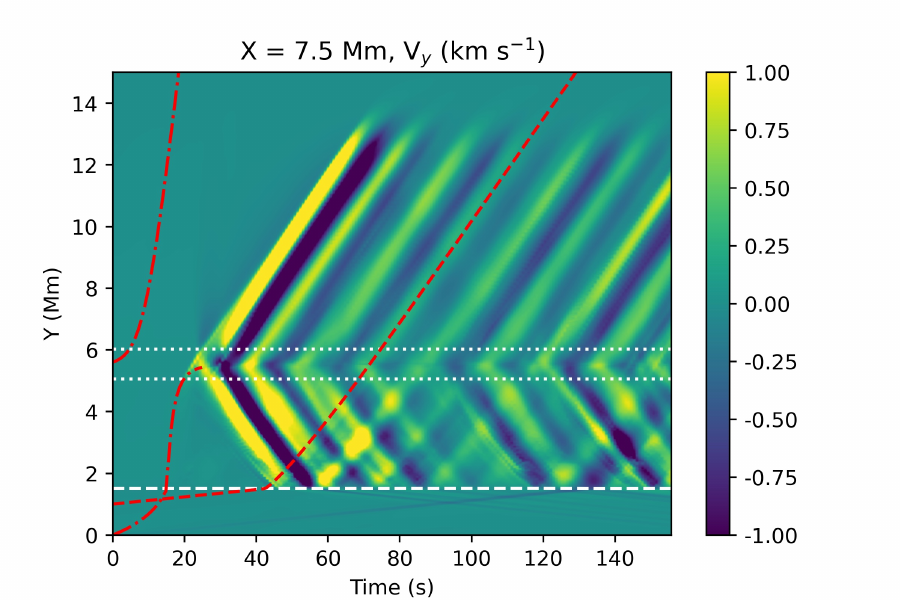}
    \includegraphics[trim={1.cm 0.cm 3.5cm 0.5cm},clip,scale=0.55]{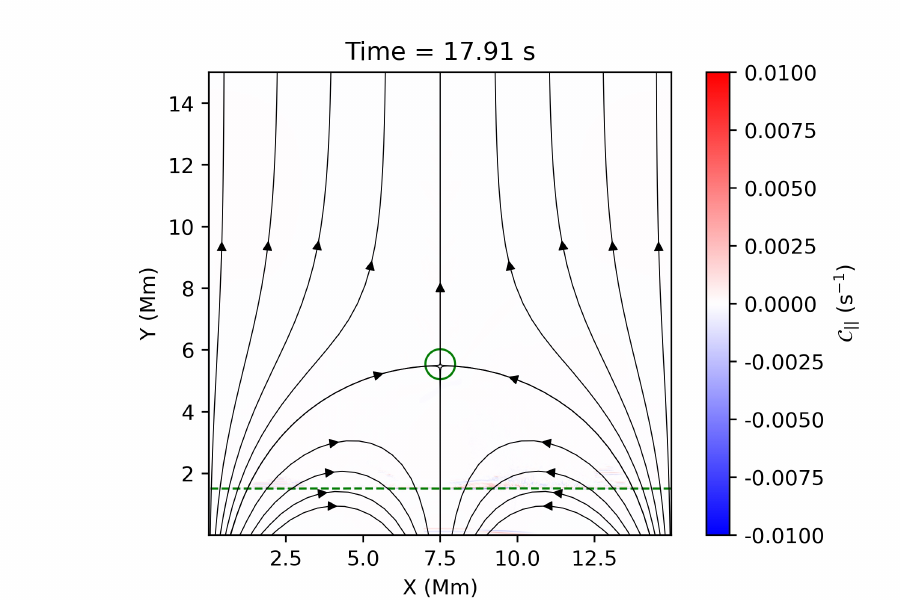}
    \includegraphics[trim={3.5cm 0.cm 0.cm 0.5cm},clip,scale=0.55]{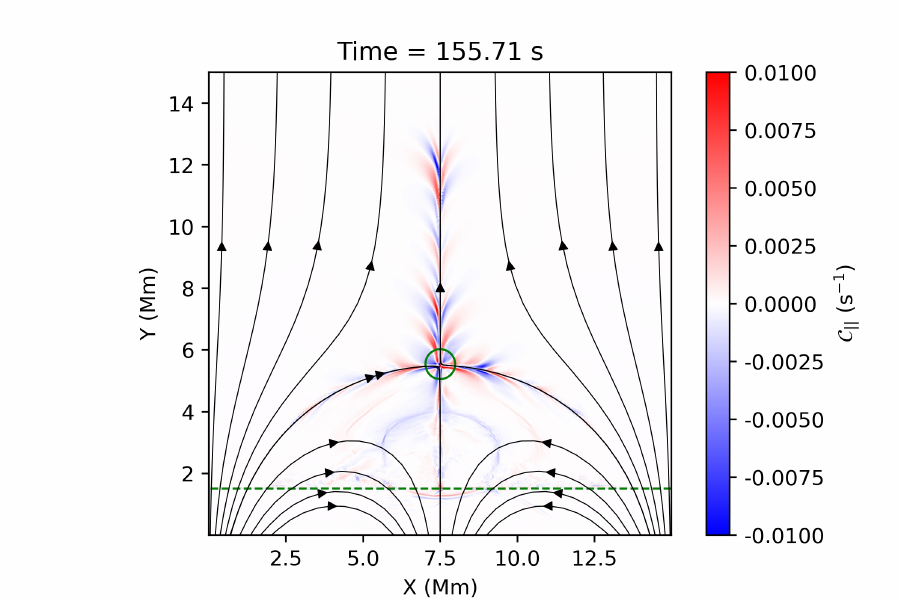}
    }
    \caption{Time-distance profiles of the $V_x$ (top left panel) and $V_y$ (bottom left panel) velocity components, along a vertical slit at $X=7.5$ Mm. The horizontal white dashed lines show the top of the transition region and the dotted white lines show the vertical borders of the $\beta=1$ layer. The red dashed lines shows the vertical trajectory over time for a sound wave, launched from $Y=0$ and the red dashed-dotted lines shows the vertical trajectory over time for an wave propagating with the Alfv\'{e}n speed, launched from $Y=0$ and from $Y=5.6$ Mm (i.e. above the null point), at $t=0$. The dual panels on the right show snapshots of the fast (top) and slow (bottom) wave identifiers, at two different times.} \label{fig:ztslits}
\end{figure*}

\begin{figure*}
    \centering
    \resizebox{\hsize}{!}{
    \includegraphics[trim={0.cm 0.cm 0.cm 0.cm},clip,scale=0.5]{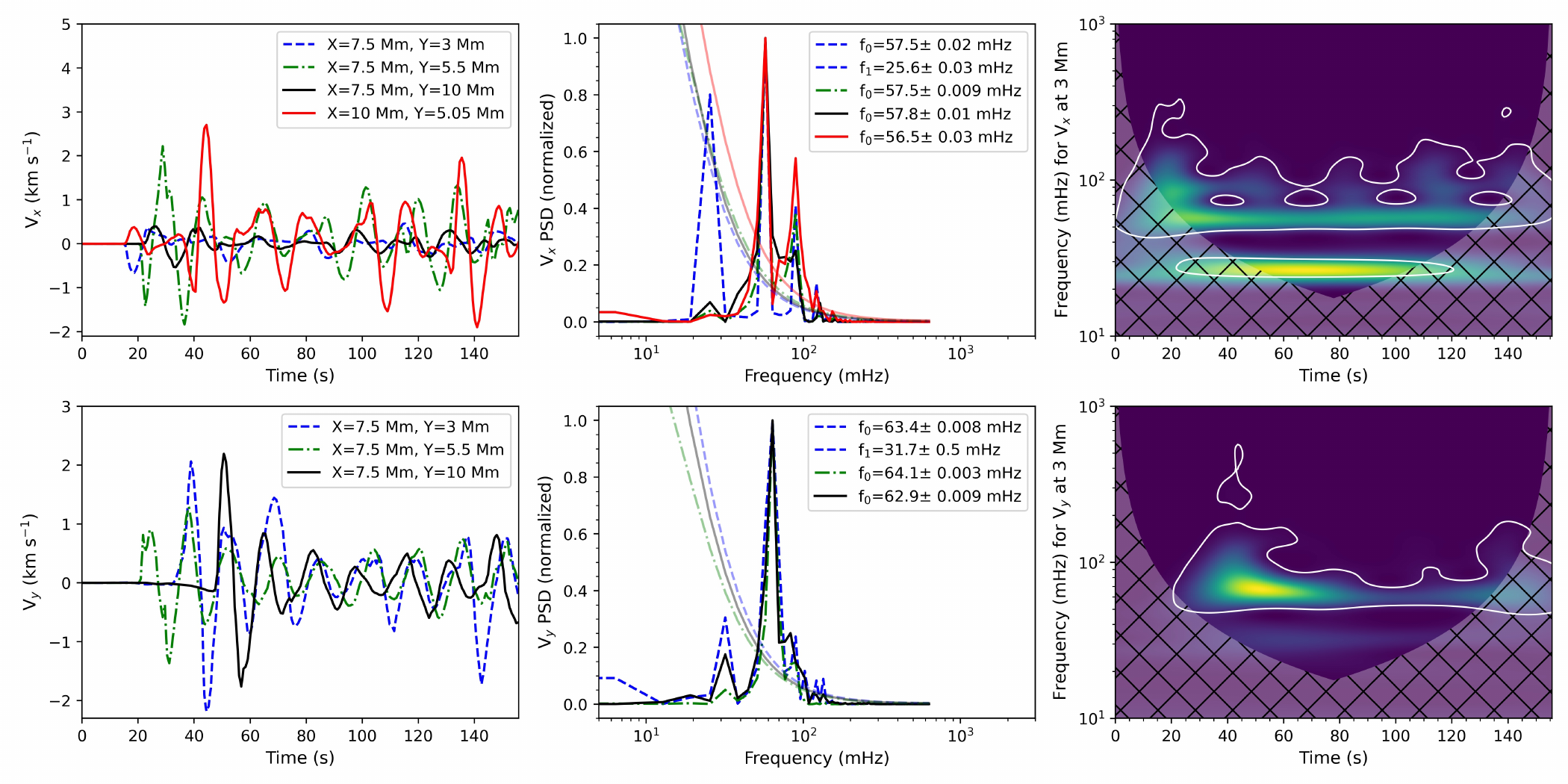}
    }
    \caption{Velocity signals at different locations (left panels) and their respective Fourier spectra (middle panels) for the $V_x$ and $V_y$ velocities. Alongside the spectra, the $95\%$ confidence levels are plotted in less opaque curves, shown here in matching colours and line styles to the respective spectra. Wavelet spectra for the $V_x$ and $V_y$ velocity signals at the $(X,Y)=(7.5,3)$ Mm point are shown on the right panels. The white contours encompassing the spectra denote the $95\%$ confidence interval. The shaded area is the cone of influence.}    \label{fig:velocityFourier}
\end{figure*}

\begin{figure*}
    \centering
    \resizebox{\hsize}{!}{
    \includegraphics[trim={0.cm 1.2cm 3.6cm 0.cm},clip,scale=0.55]{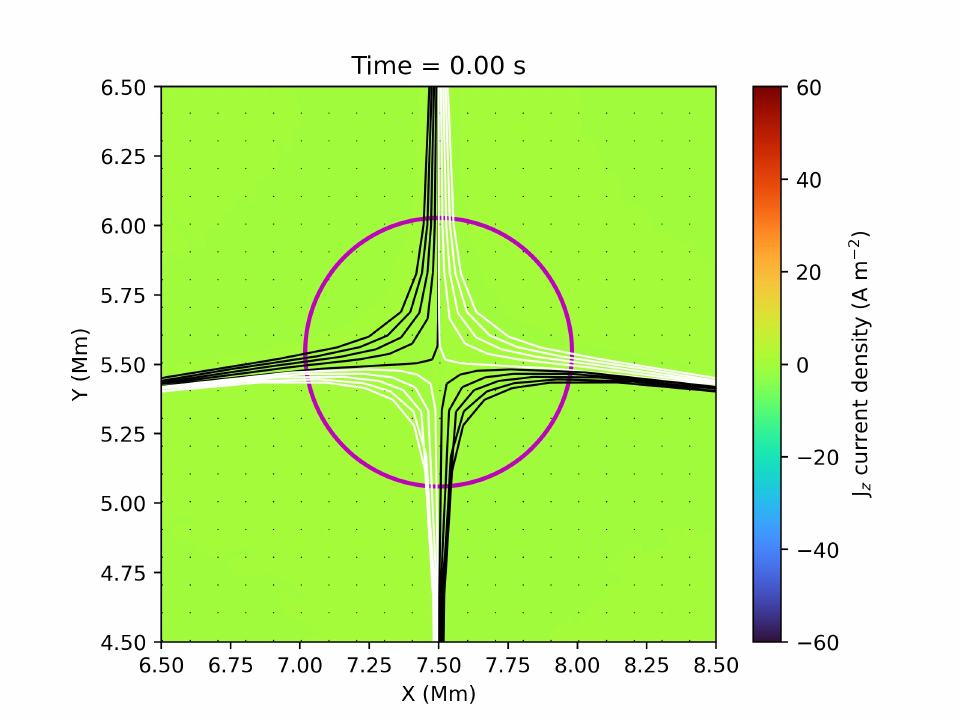}
    \includegraphics[trim={2.5cm 1.2cm 3.6cm 0.cm},clip,scale=0.55]{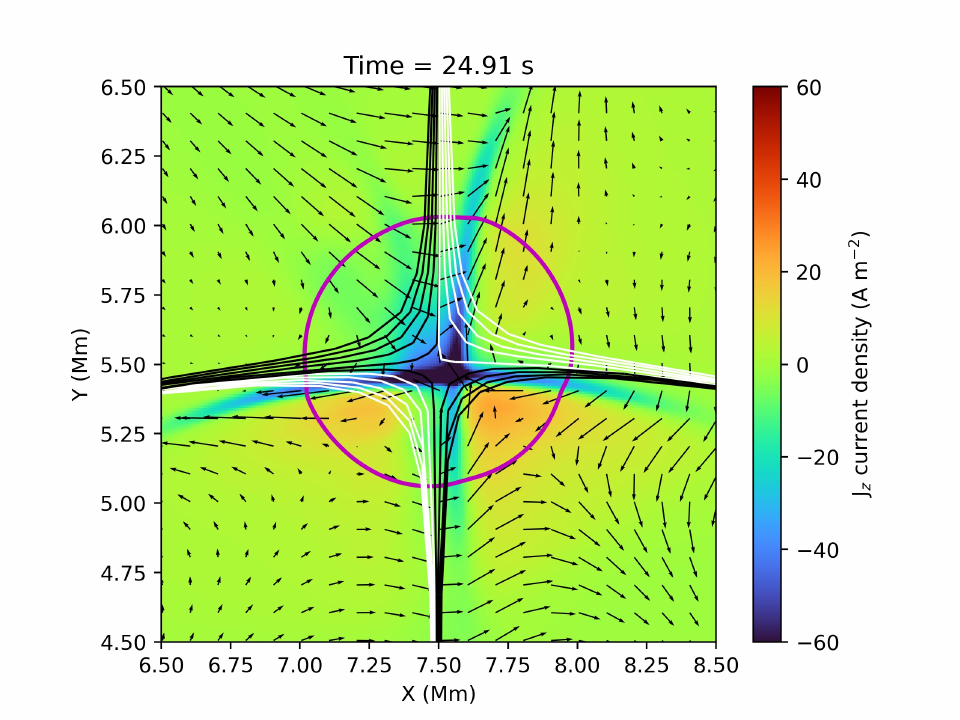}
    \includegraphics[trim={2.5cm 1.2cm 3.6cm 0.cm},clip,scale=0.55]{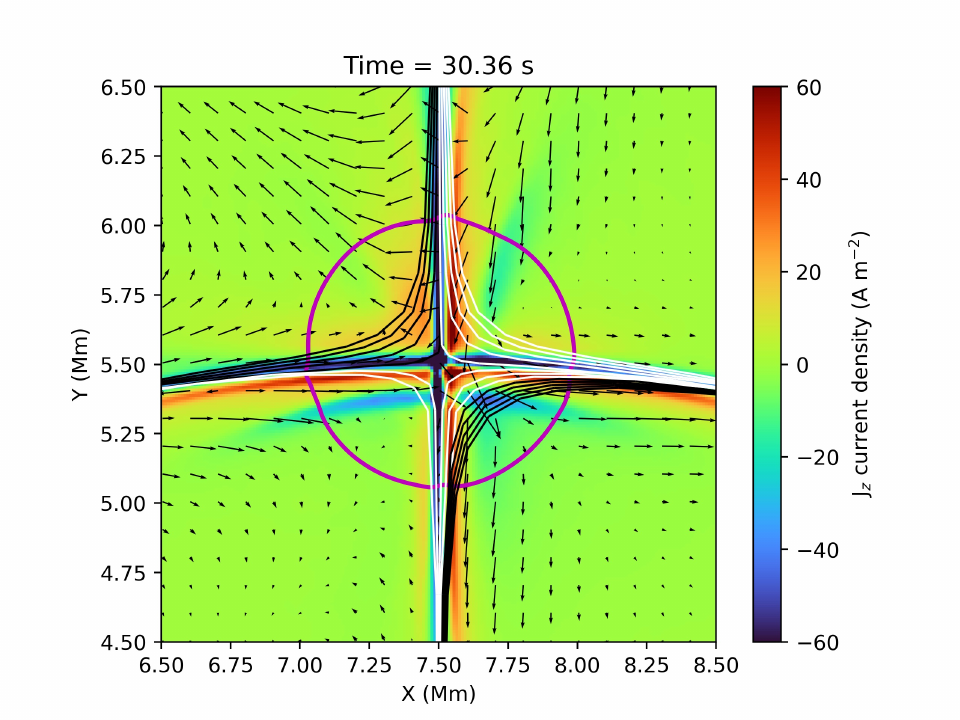}
    \includegraphics[trim={2.5cm 1.2cm 3.6cm 0.cm},clip,scale=0.55]{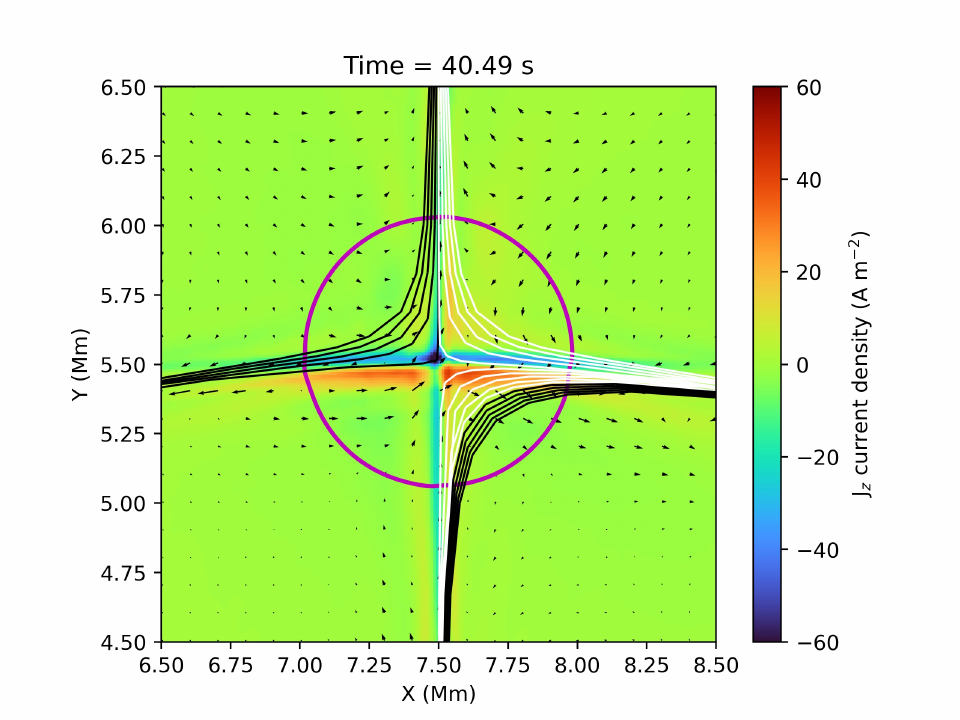}
    \includegraphics[trim={2.5cm 1.2cm 0.cm 0.cm},clip,scale=0.55]{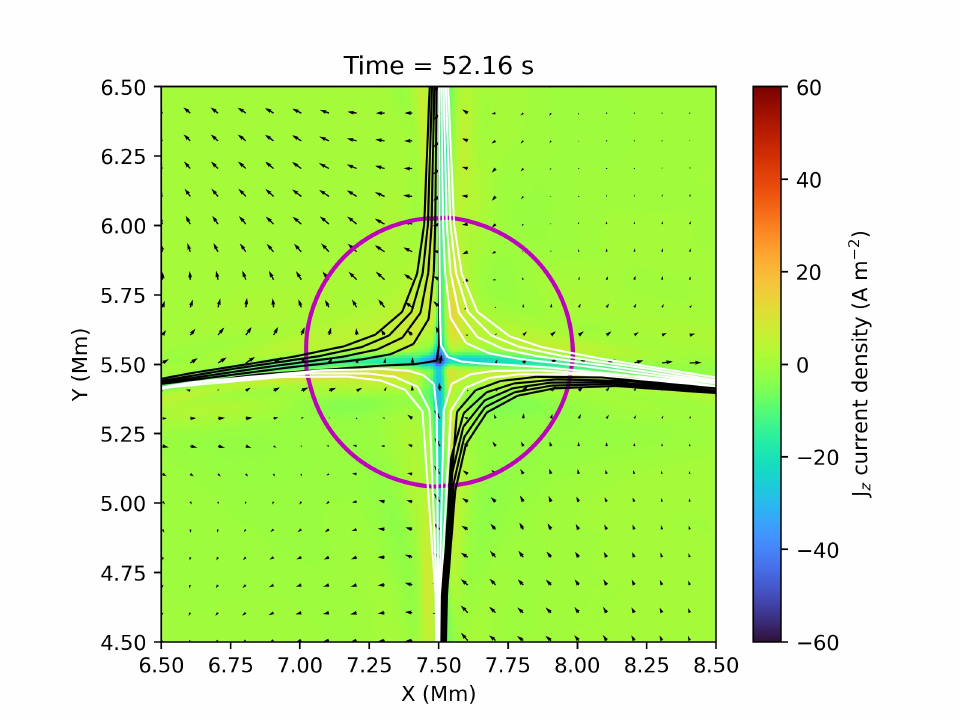}
    }
    \resizebox{\hsize}{!}{
    \includegraphics[trim={0.cm 0.cm 3.6cm 0.cm},clip,scale=0.55]{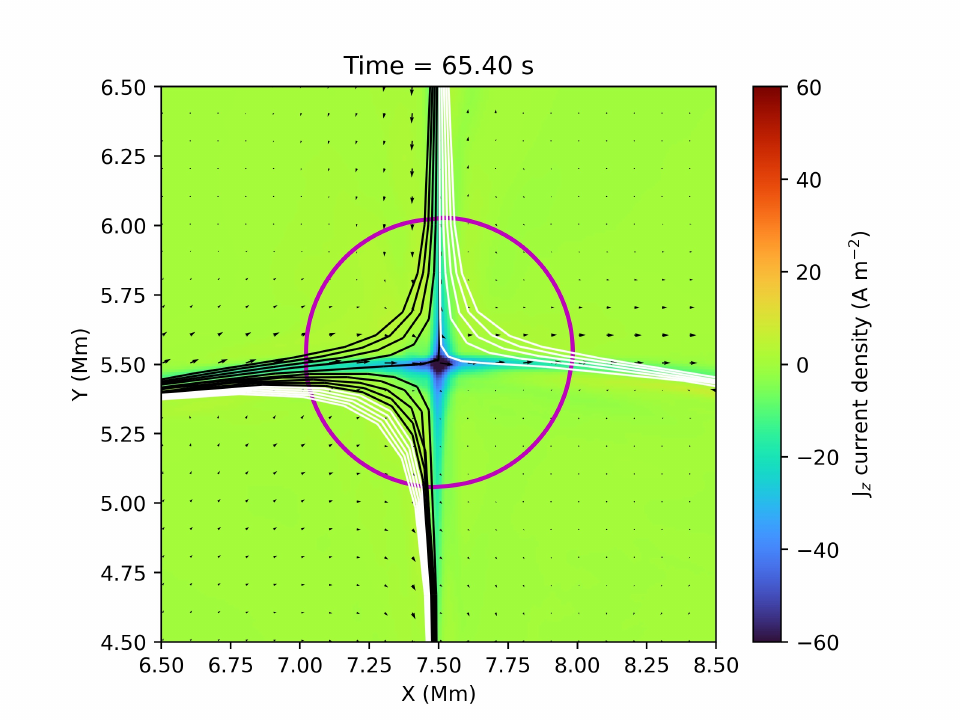}
    \includegraphics[trim={2.5cm 0.cm 3.6cm 0.cm},clip,scale=0.55]{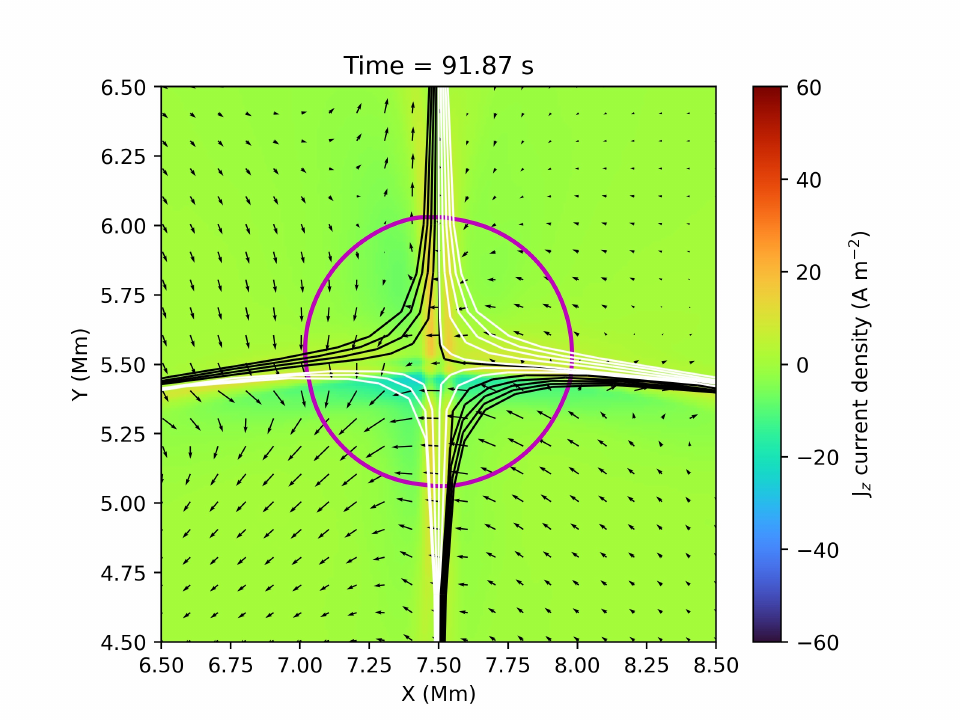}
    \includegraphics[trim={2.5cm 0.cm 3.6cm 0.cm},clip,scale=0.55]{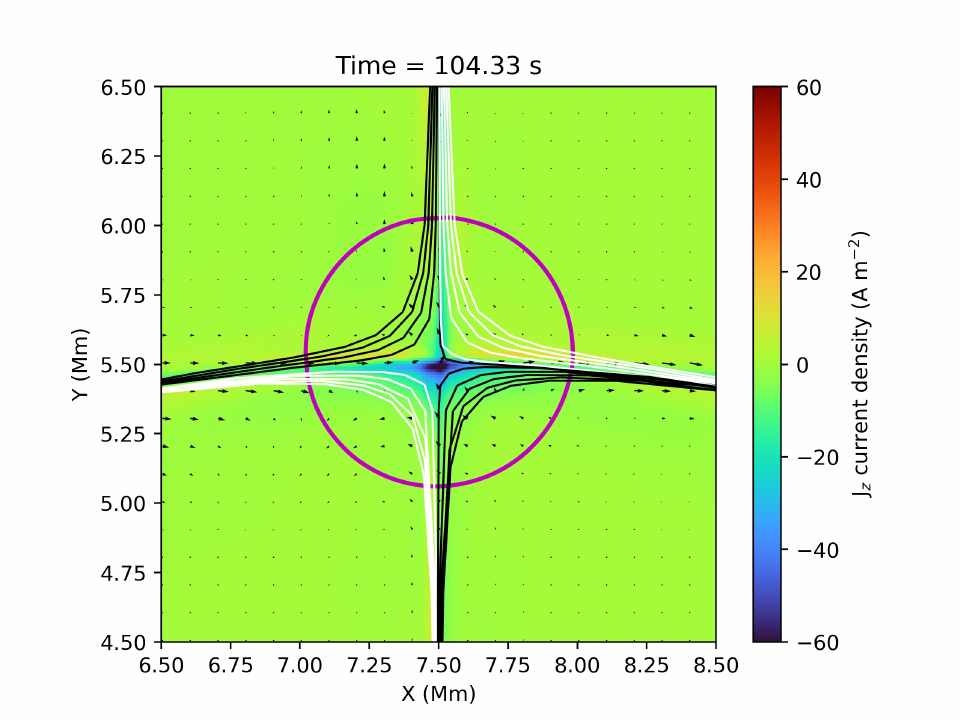}
    \includegraphics[trim={2.5cm 0.cm 3.6cm 0.cm},clip,scale=0.55]{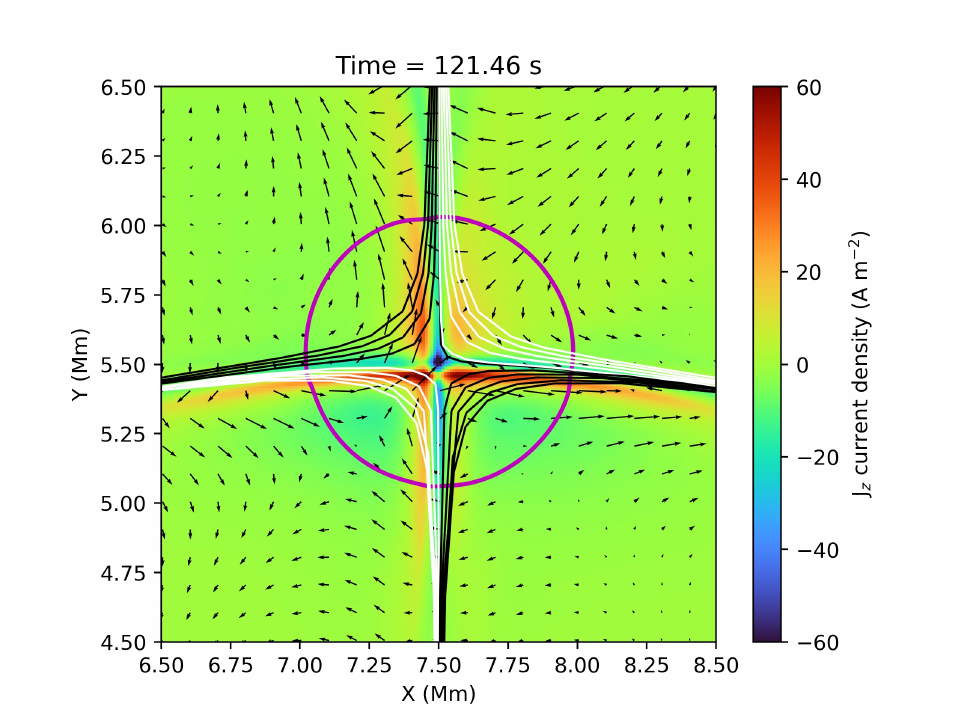}
    \includegraphics[trim={2.5cm 0.cm 0.cm 0.cm},clip,scale=0.55]{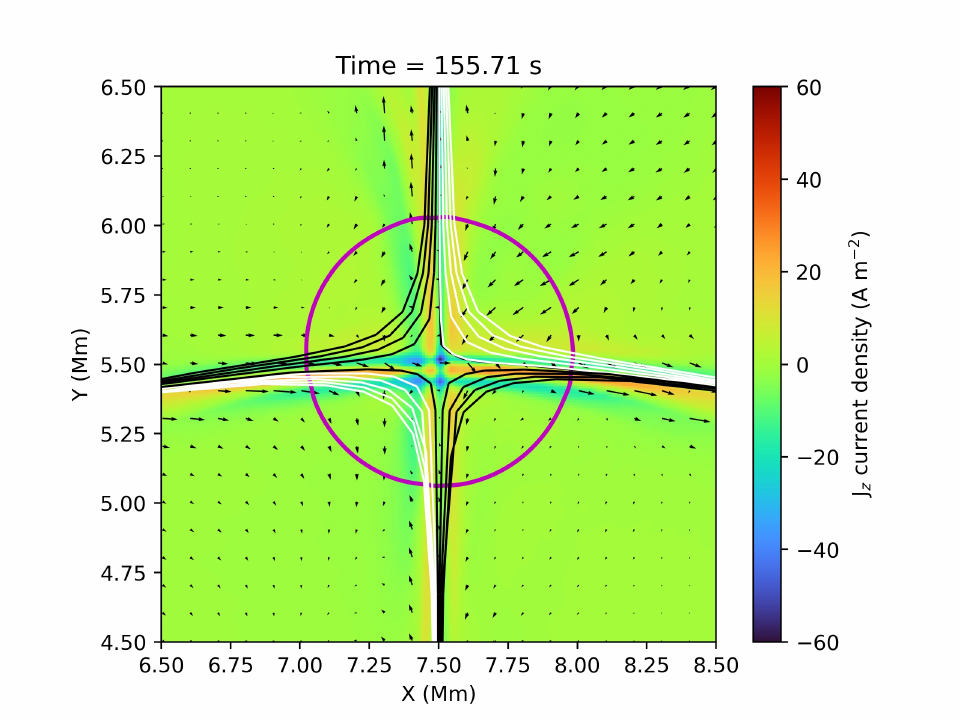}
    }
    \caption{Close-up snapshots of the $J_z$ current density at the null point during the reconnection events. Also shown here are the $\beta=1$ layer (purple circle), a sample of the magnetic field lines (in black and white) the vectors for the velocity field. An accompanying animation can be found in the online version of this manuscript.}    \label{fig:nullpointfieldlines}
\end{figure*}

\begin{figure*}
    \centering
    \resizebox{\hsize}{!}{
    \includegraphics[trim={0.cm 0.cm 0.cm 0.cm},clip,scale=0.5]{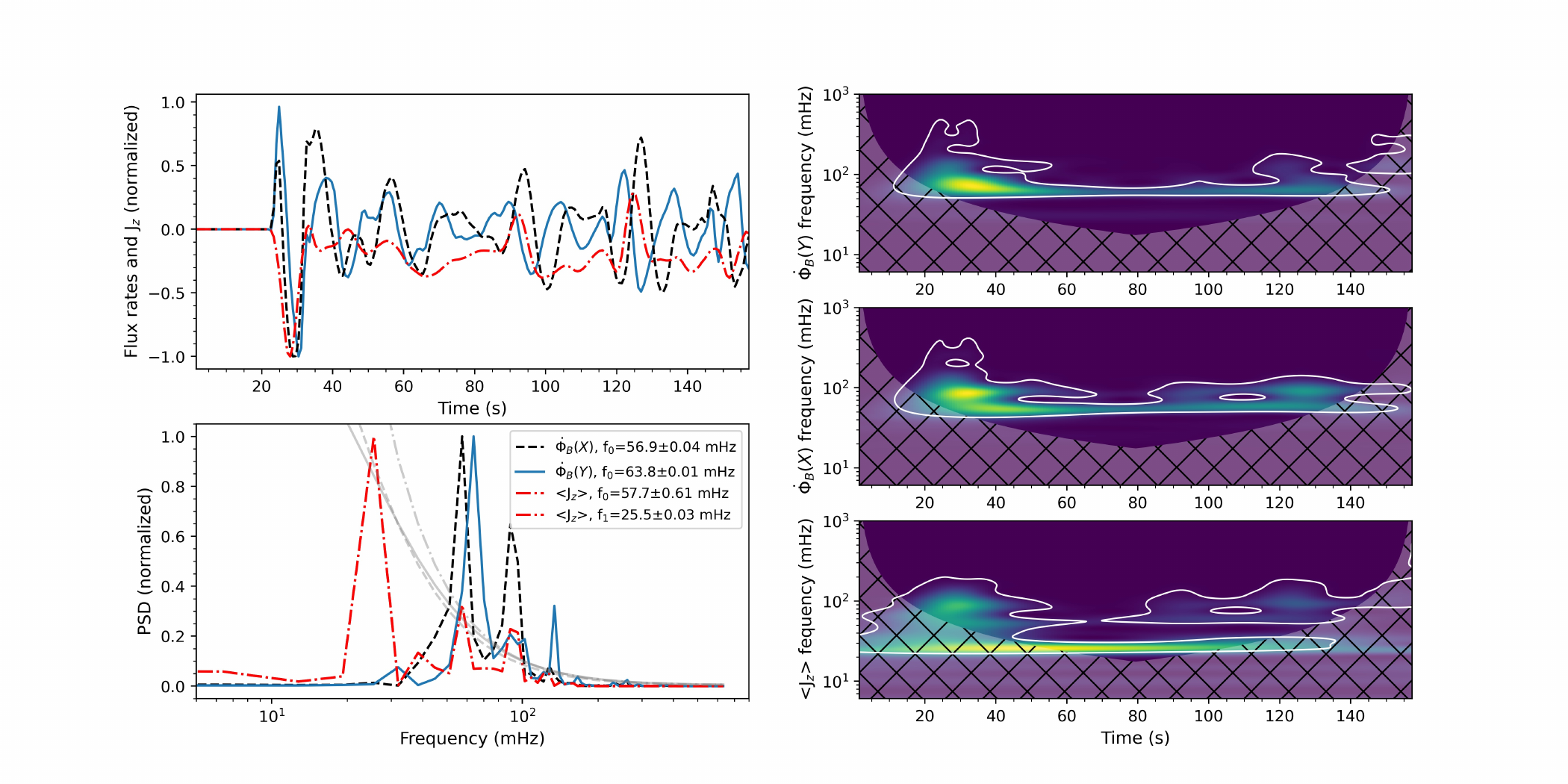}}
    \caption{Time series of the average $J_z$ current density at the null, as well as the magnetic flux rates $\dot{\Phi}_B(X,Y)$ for the $B_{x,y}$ magnetic field components, near the null point. Shown here are also their respective Fourier, with the $95\%$ confidence levels in grey and matching line styles, and wavelet spectra. The white contours encompassing the wavelet spectra denote the $95\%$ confidence interval and the shaded area in each one is the cone of influence.} \label{fig:nullpointreconnection}
\end{figure*}

\begin{figure*}
    \centering
    \resizebox{\hsize}{!}{
    \includegraphics[trim={0.5cm 0.cm 2.cm 0.cm},clip,scale=0.5]{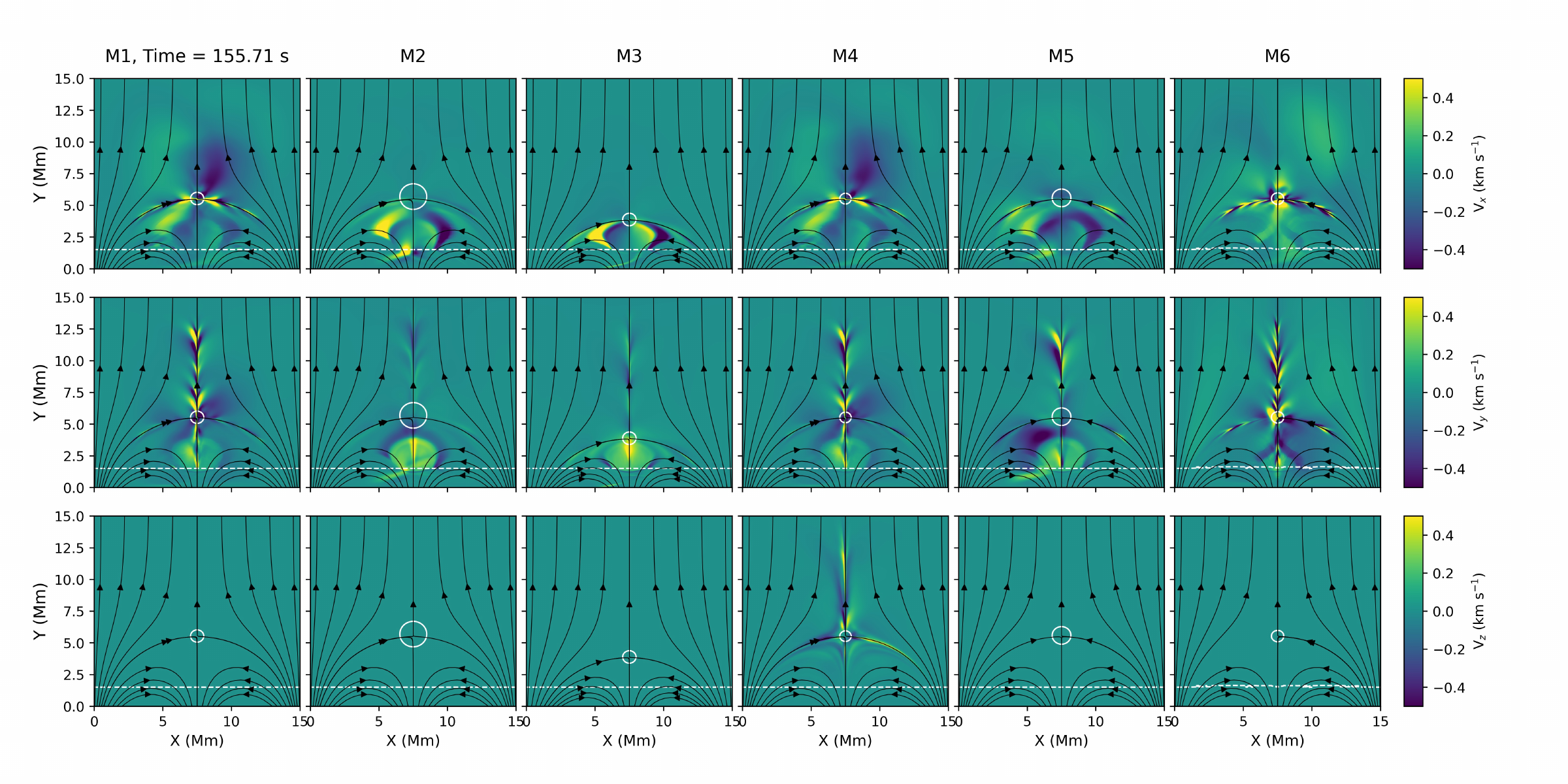}}
    \caption{Profiles of the $V_x$ (top row), $V_y$ (middle row) and $V_z$ (bottom row) velocities for each model M1 to M6 (from left to right) at $t=155.71$ s. Also shown here are the top of the transition region (white dashed horizontal line) and the $\beta=1$ layer (white contour line).} \label{fig:comparative}
\end{figure*}

\begin{figure*}
    \centering
    \resizebox{\hsize}{!}{
    \includegraphics[trim={0.cm 0.cm 0.3cm 0.cm},clip,scale=0.3]{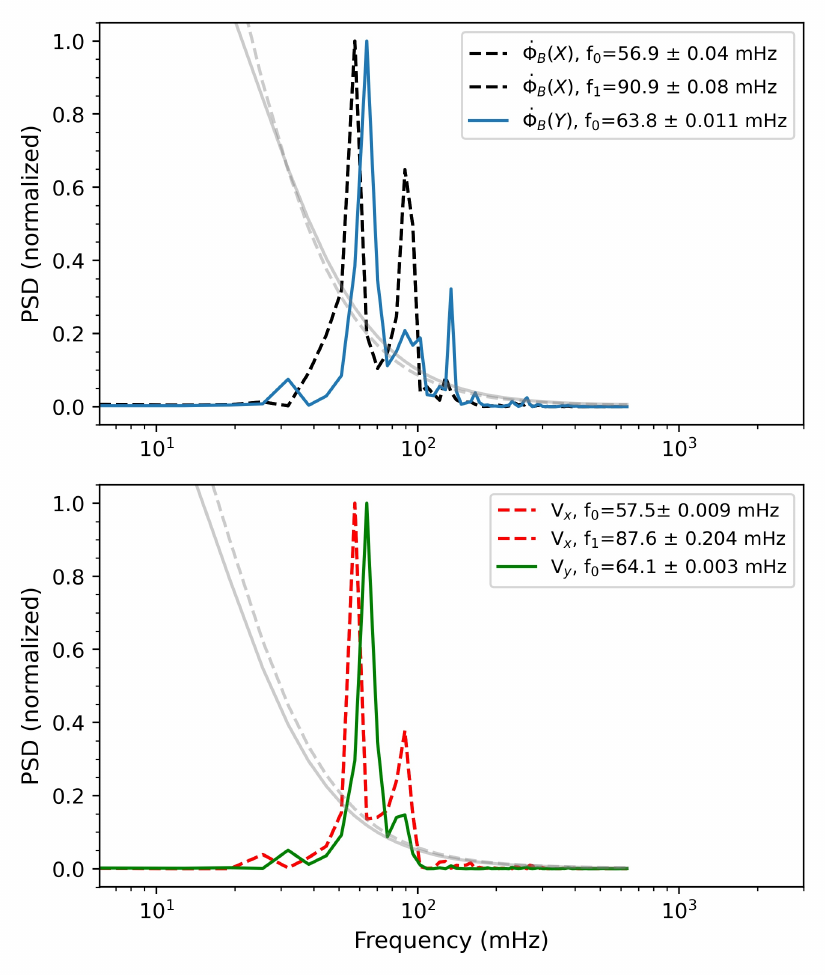}
    \includegraphics[trim={1.5cm 0.cm 0.3cm 0.cm},clip,scale=0.3]{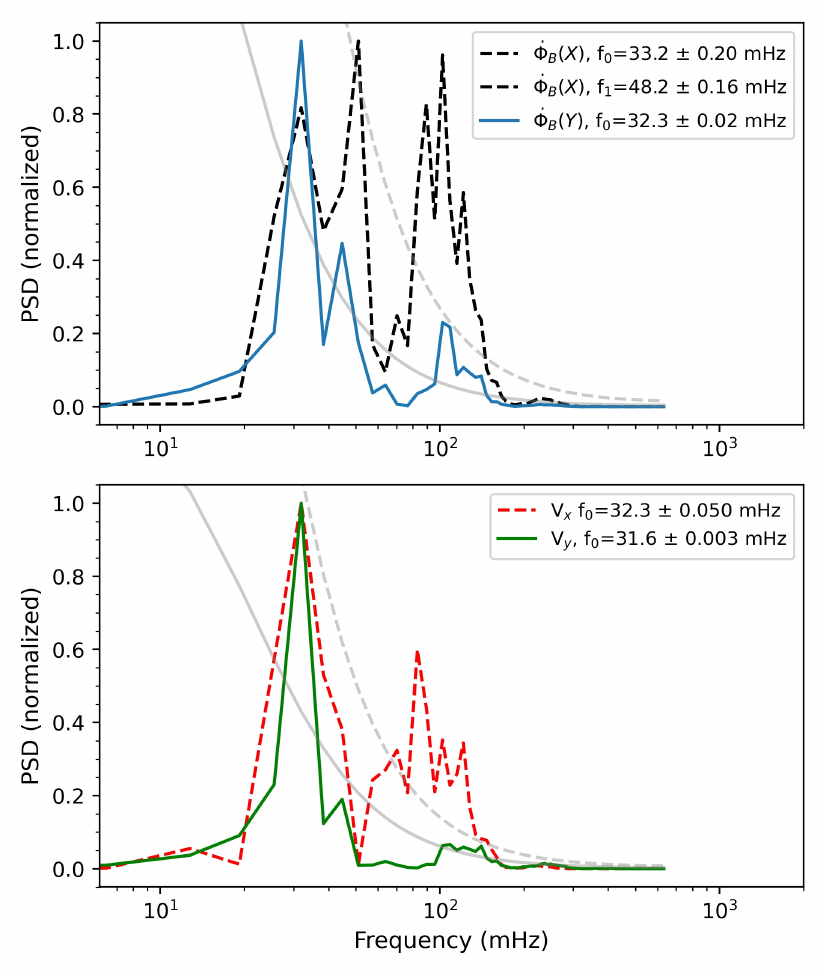}
    \includegraphics[trim={1.5cm 0.cm 0.3cm 0.cm},clip,scale=0.3]{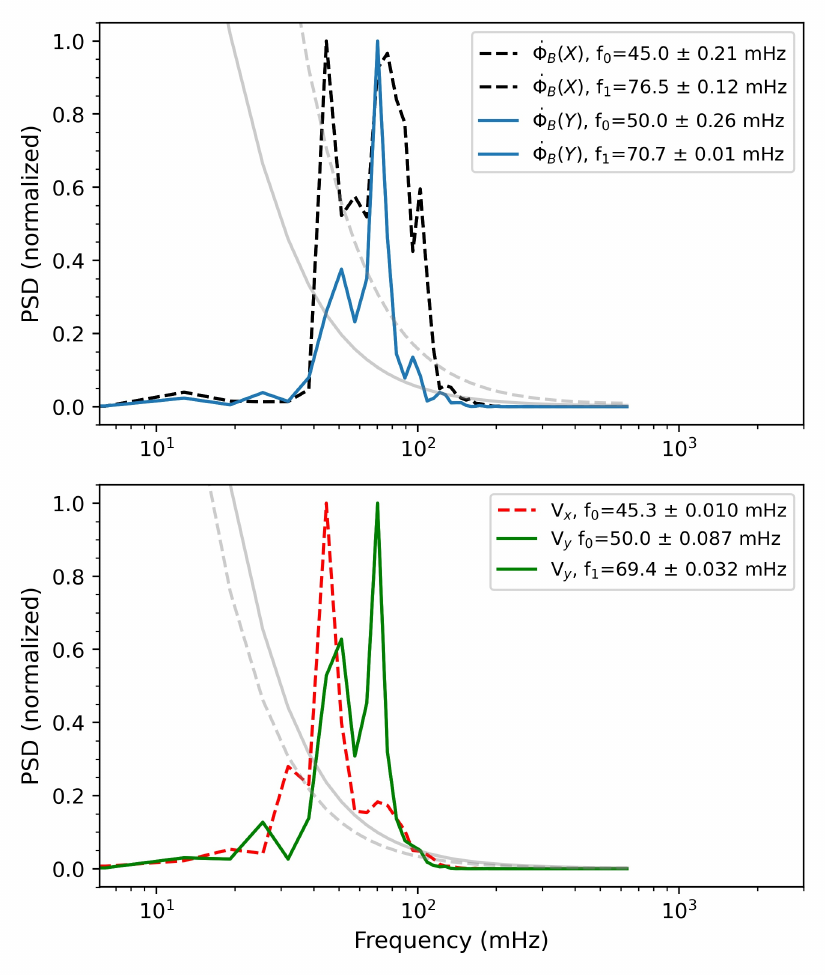}
    }
    \resizebox{\hsize}{!}{
    \includegraphics[trim={0.cm 0.cm 0.3cm 0.cm},clip,scale=0.3]{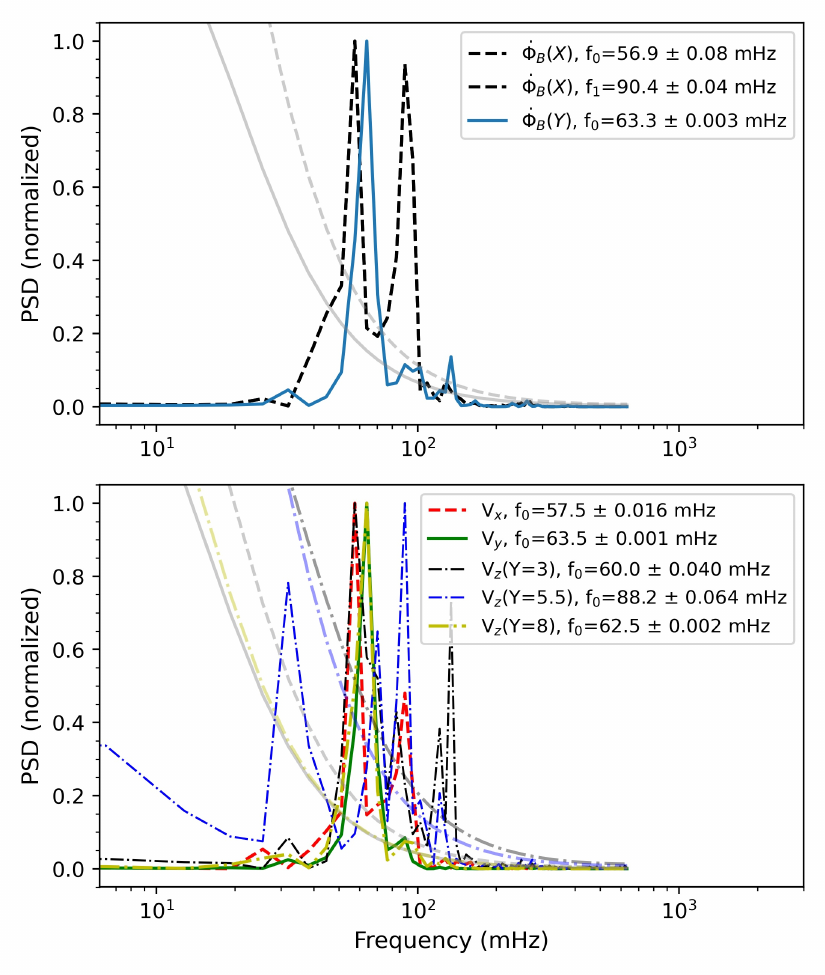}
    \includegraphics[trim={1.5cm 0.cm 0.3cm 0.cm},clip,scale=0.3]{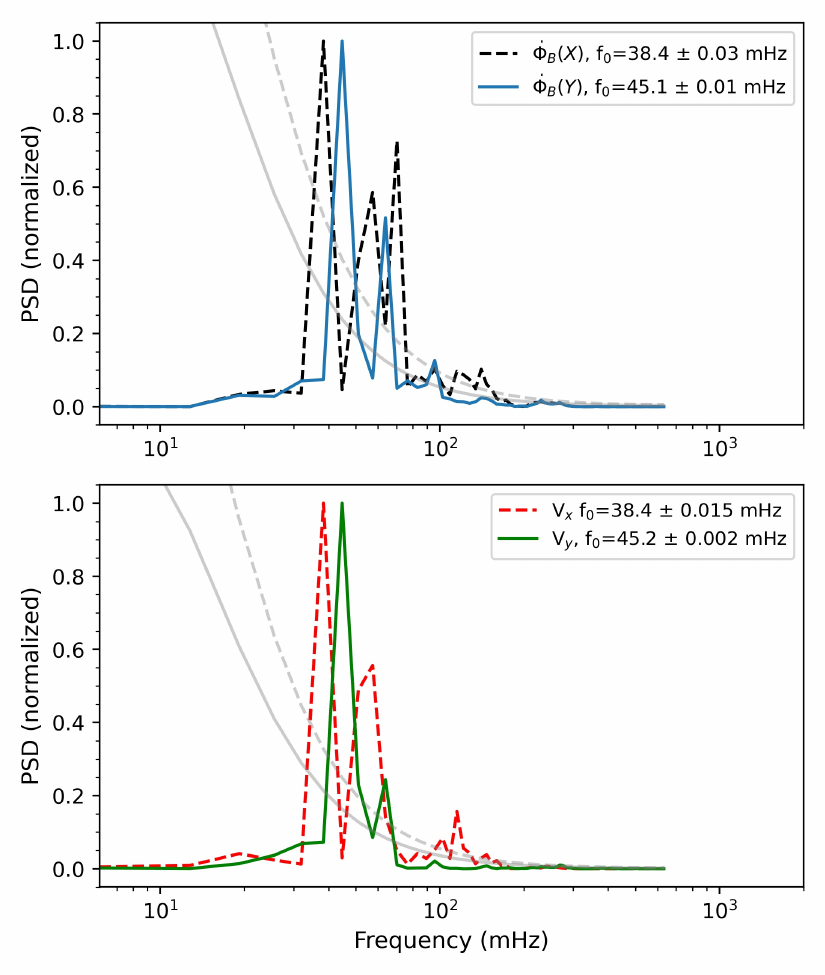}
    \includegraphics[trim={1.5cm 0.cm 0.3cm 0.cm},clip,scale=0.3]{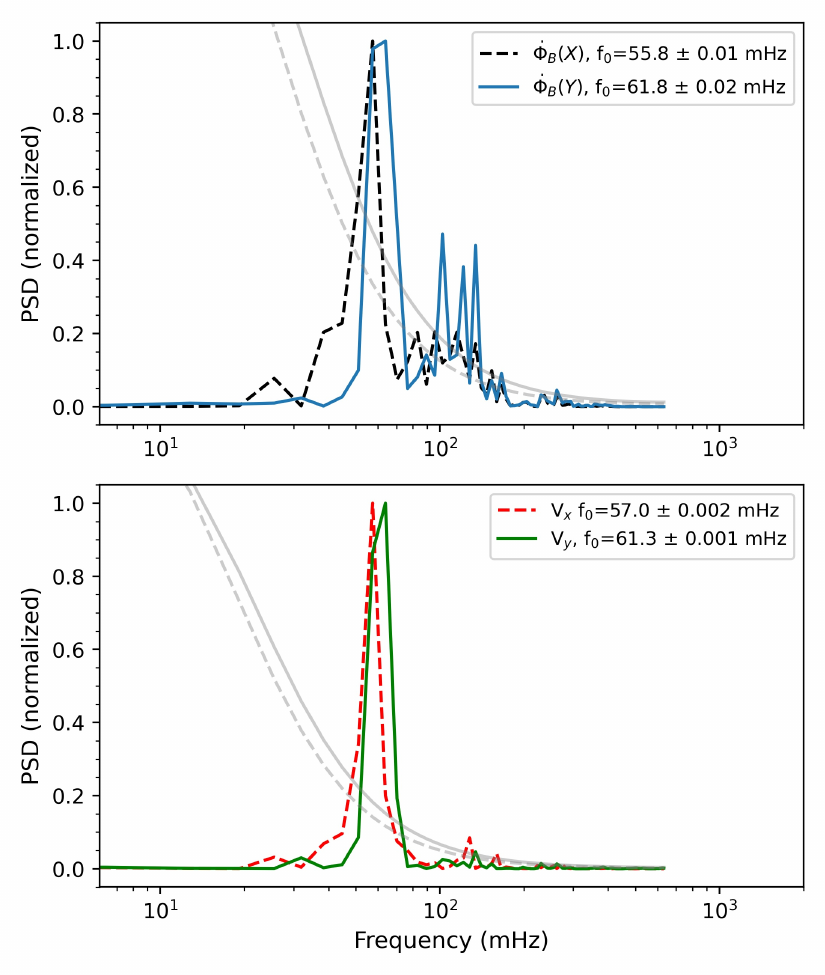}
    }
    \caption{Fourier spectra of the magnetic flux rates $\dot{\Phi}_B(X,Y)$ and of the $V_{x,y}$ velocity components at the location of the null points, for models M1, M2 and M3 (top three dual panels) and for M4, M5 and M6 (bottom three dual panels). The $95\%$ confidence level curves are shown for each quantity in less opaque, matching style grey lines, or matching colored lines for the $V_z$ signals of model M4. } \label{fig:parametercavity}
\end{figure*}

Our numerical setup consists of a 2D null point consisting of a potential magnetic field embedded in a gravitationally stratified atmosphere, inspired by the model of \citet{Santamaria2015}. We solved the compressible MHD equations for a hydrogen plasma using the PLUTO code \citep{mignonePLUTO2007}, performing a parameter study of setups with slightly different initial conditions. For all simulations we used a third-order Runge-Kutta method to advance the time step and a third-order piecewise parabolic spatial reconstruction method, while the fluxes were calculated with the HLL (Harten, Lax, van Leer) approximate Riemann solver. The $\nabla\cdot\mathbf B=0 $ condition is met via the constrained transport method. No explicit resistivity is present in our model; however, the finite discretisation gives rise to numerical effective resistivity, many orders of magnitude higher than the expected value in the solar atmosphere.  During our past studies of waves and reconnection in coronal plasma \citep[e.g.][]{Karampelas2023ApJ...943..131K} we have reached a range of values for the effective Magnetic Reynolds number of $R_m \in [10^4-10^6]$. These values are far lower than the values expected in the solar corona \citep[$\sim 10^8-10^{12}$, see][]{HoodHughes2011PEPI..187...78H} and the only way to reach such values numerically is through extremely costly high-resolution simulations. No thermal conduction terms were introduced in this setup, while explicit viscosity was only present in the initial stages of the relaxation phase of our simulations (for $\sim 155.8$ s), before the velocity perturbation is introduced through the driver. During that part of the relaxation phase, the shear and bulk viscous transport coefficients were equal to $0.001$ kg m$^{-1}$ s$^{-1}$, while both were kept equal to zero for the rest of the each simulation. 

Our 2D Cartesian domain has dimensions $X\in [0,\,15]$ Mm and $Y\in [0,\,20]$ Mm and a resolution of $(X,Y) = (1500,2000)$ points, with $\delta X = \delta Y = 10$ km. We used periodic boundary conditions in the $X$ direction and open, outflow conditions at the $Y=20$ Mm boundary. At the bottom boundary $(Y=0)$ we used zero-gradient conditions for the three components of the magnetic field. This prevents the generation of additional currents near the bottom boundary, where the magnetic field is the strongest, aiding in code stability.
We also use symmetric conditions for the density and pressure and antisymmetric conditions for the $X,Z$ components of the velocity, to prevent plasma evacuating through the bottom boundary due to gravity. For the $Y$ component of the velocity, we used antisymmetric conditions during the relaxation phase of each simulation and for $t>7.79$ s. We note here that in our results, we set $t=0$ as the time when we start driving our system from the bottom boundary. Therefore, the relaxation phase takes place before $t=0$. For $0\leq t \leq 7.79$ s, the $y$-component velocity  at the bottom boundary is equal to:
\begin{align}
    &V_y = 500\, \sin(2\pi t / P)\, e^{-(X-10)^2}, \, \text{in m s}^{-1}.
\end{align}
This introduces a sinusoidal velocity pulse, driven only for half a period ($\Delta t = 7.79$ s for a period of $P = 15.58$ s). 

Alongside the periodic boundaries in the $x$ direction and the outflow conditions at $Y=20$ Mm, we also employ velocity rewrite layers that remove energy from the velocity perturbations, preventing any reflections or inflows from the sides and top boundaries to return to the areas or interest. In these layers, at each time-step, we divide the velocity components $V_i$,  where $i=(x, y, z)$, by a coefficient $n_d$. In total we have three different layers defined by $n_d$:
\begin{align}
    &n_d = 1.001 + 0.001\tanh(Y - 14), \, Y>13 \text{ Mm,}\\
    &n_d = 1.001 + 0.001\tanh(X - 13), \, Y<15 \text{ and } X>12 \text{ Mm}\\
    &n_d = 1.001 + 0.001\tanh(2 - X), \, Y<15 \text{ and } X<3 \text{ Mm.}
\end{align}
The above equations define two overlap regions ($13<Y<15$ Mm, $X < 3$ Mm and  $13<Y<15$ Mm, $X > 12$ Mm) where two different dissipation schemes are active. In these regions we allow the code to apply two dissipation layers simultaneously, after we made certain that this would not affect the domain near the null point, which is our area of interest.

\subsection*{Default atmosphere and magnetic field}
We initialized our background atmosphere using a 2D model in hydrostatic equilibrium along the vertical $Y$ direction, using a uniform gravity ($g_y(Y) = 274$ m s$^{-2}$). For the mean atomic weight $\mu$ we considered:
\begin{align}
    &\mu = 0.5\, (1.5 - 0.5\,\tanh((Y-1.5)100)),
\end{align}
taking $\mu = 1$ at the chromospheric part of our atmosphere and $\mu = 0.5$ in the corona where the plasma is fully ionized, similarly to \citet{Santamaria2015}. We use a temperature profile, inspired by \citet{AschwandenSchrijver2002}:
\begin{align}
&T(Y\geq 1.5 \text{ Mm}) = T_\mathrm{Ch} + (T_\mathrm{C} - T_\mathrm{Ch})\left( 1 - \left[ \frac{10-Y}{10-1.5} \right]^2 \right)^{0.3},\\
&T(Y< 1.5 \text{ Mm}) = T_\mathrm{Ch},
\end{align}
where $1.5$\,Mm is the width of our chromosphere, $T_\mathrm{Ch} = 0.01$\,MK is the temperature of the chromosphere and $T_\mathrm{C} = 1$ MK is the temperature at $Y=10$ Mm. The density at the bottom of the chromosphere ($Y=0$) is equal to $\rho_\mathrm{Ch} = 2.9 \times 10^{-8}$ kg m$^{-3}$. The value of the pressure at $Y=0$ was calculated via the ideal gas law. Finally, we solved the equations of the hydrostatic equilibrium along the $Y$ direction using a forward Euler method, to construct the vertical profiles of the density and pressure. The end result is an atmospheric profile with a sharp transition region between the chromospheric and the coronal part of our domain. This transition region is not properly resolved by the resolution in our domain. However, in this study we are not studying the thermodynamic response of our system, or the mass flux across along height. Instead, we use the transition region as a semi-elastic wall, due to the strong density gradient, that will affect the propagation of waves across the domain.

For the magnetic field, we considered a potential magnetic field:
\begin{align}
    & B_x(X,Y) = B_0 \exp(-k\,Y)\sin(k\,X),\\
    & B_y(X,Y) = B_u + B_0 \exp(-k\,Y)\cos(k\,X), \text{ with } k=2\pi/15,
\end{align}
where $B_0$ and $B_u$ are the magnetic field strengths in the chromosphere and corona, respectively and $k$ defines the horizontal and vertical spatial scales. For the default setup, we take $B_0=100$ G and $B_u=10$ G. Our magnetic field is composed by two vertical flux tubes at the edges of our domain, separated by an arcade-shaped magnetic field, with a magnetic null point (here, an X-point) located at $X=7.5$. The height ($Y$) of the null point depends on the model considered at each case (see following subsection). For the default configuration, we have $Y=5.55$ Mm.

Although our magnetic field is a potential one, due to the finite discretization of our domain weak non-force free field components and weak currents are generated. In order to allow for the setup to reach a semi-equilibrium state, we perform a relaxation for a total time of $\Delta t_r$ (different for each setup), as mentioned earlier. During the initial stages of the relaxation phase, we use a uniform non-zero explicit viscosity for the first $155.8$ s, as stated above. The total relaxation time for the default setup (and all others except one, see below) is $\Delta t_r = 311.6$ s. The profiles of the temperature, density, $B_x$ and $B_y$ magnetic field, post relaxation are shown in Figure \ref{fig:inicon}. In these panels, the magnetic field lines are also shown, alongside the plasma $\beta = 1$ layer around the null point (black circle) and the top of the transition region ($T=0.1$ MK, black dashed line).

\subsection*{Alternative magnetic field and plasma configurations}
Apart from the default setup described above, in this study we are also considering different configurations of the hydrostatic model and magnetic field used. The different configurations are shown in Table \ref{table:1}. Model M1 is the default model described above, while models M2 and M3 have the same background atmosphere but a modified magnetic field configuration (i.e. different planar field strength and/or null point location). Model M4 is the same as M1 apart from a non-zero $B_z$ component of a guide magnetic field in the $z$ direction. The planar components of the magnetic field in M4 still exhibit an X-point configuration, but the latter is now not a true null point, due to the non-zero $B_z$. It is instead a configuration of an X-line. M5 is again the same as M1, but the density at the bottom of the chromosphere is now double the amount of than in M1, for the same temperature profile. This leads to a heavier atmosphere that in the M1 model. Finally, M6 is a setup visually resembling a pseudostreamer (but without any outward flow). Model M6 has the same initial magnetic field and chromospheric temperature and density as M1, but a transverse variation on the temperature and density. To construct M6, we changed the prescribed coronal temperature profile, before solving the equations of the hydrostatic equilibrium. In particular, we calculated the equation of the magnetic field vector potential:
\begin{equation}
    \mathbf{A} = A_z\,\hat{k} = -B_uX\,\hat{k}-\frac{B_0}{k}e^{-kY}\sin(kX)\, \hat{k}, \text{with } k = \frac{2\pi}{15}
\end{equation}
and we solved for the height $Y_B(A_z,X)$. We then iterated along the $X,Y$ directions of the domain, looping for different values of the magnetic potential (here, $A_z \in \left[ -65, -40 \right]$) and for $Y_b\sim Y$ we incrementally increased the local value of the temperature. At the end of the iteration we normalise the temperature, so that its maximum value is equal to the maximum value of the M1 model, while we keep the initial temperature at the chromosphere at $T_{ch} = 0.01$\,MK. Finally, after calculating the density and pressure, we allow our model to relax for twice as long as the other models, or $\Delta t_r = 623.2$\,s for the M6 model. 
The 2D plots of temperature and density for the 2D null point atmosphere for models M2, M3, M5 and M6 are shown in Figure \ref{fig:alternative}, alongside the plasma $\beta=1$ layer around the null point (black circle) and the top of the transition region ($T=0.1$ MK, purple dashed line). The temperature and width of the chromosphere is the same for all models, while the height of the resulting null point, strength of the mean magnetic field, as well as density and temperature of the corona vary by model (see Table \ref{table:1}).

\begin{table}
\caption{List of setups.}              
\label{table:1}      
\centering                                      
\begin{tabular}{l c c c c c}          
\hline\hline                        
Model & $\rho_\mathrm{Ch}$ (kg m$^{-3}$) & $B_0$ (G) & $B_u$ (G) & $B_z$ (G) & $\Delta t_r$ (s) \\    
\hline                                   
    M1 & 2.9 10$^{-7}$ & 100 & 10 & 0 & 311.6 \\      
    M2 & 2.9 10$^{-7}$ & 50 & 5 & 0 & 311.6 \\
    M3 & 2.9 10$^{-7}$ & 50 & 10 & 0 & 311.6 \\
    M4 & 2.9 10$^{-7}$ & 100 & 10 & 1 & 311.6 \\
    M5 & 5.8 10$^{-7}$ & 100 & 10 & 0 & 311.6 \\
    M6 & 2.9 10$^{-7}$ & 100 & 10 & 0 & 623.2 \\
\hline                                             
\end{tabular}
\tablefoot{Initial conditions for the chromospheric density ($\rho_\mathrm{Ch}$), magnetic field ($B_u,\, B_0,\, B_z$) and total relaxation time ($\Delta t_r$) for the six different models used in this study.}
\end{table}

\section{Results} \label{sec:results}
We run the simulations for each of our models (M1 to M6) for a total of $\Delta t = 156$\,s per case, corresponding to $10$ driving wave periods, after the initial MHD relaxation process. In all of our cases, plasma $\beta<1$ everywhere away from the null point, as shown in Figure \ref{fig:alternative}. 

\subsection{Periodicity of resonant cavity and reconnection}
We will start our analysis looking at first into the results of the model M1. The initial $V_y$ pulse creates a velocity perturbation that starts propagating along and across the field lines, as shown in Figure \ref{fig:2Dvelocities} from the $V_y$ and $V_x$ velocity components. Our pulse has mixed properties, in the sense that we do not explicitly excite a fast or slow MHD wave, rather the initial perturbation excites both types of wave perturbations. As shown in the same figure and the accompanying animations, as the pulse approaches the null point, it starts refracting around it, which has been seen in the past for fast wave pulses \citep[e.g.][]{McLaughlin2006b, Santamaria2018} and for mixed properties pulses \citep{Tarr2017ApJ...837...94T}. As we are going to see below in the analysis, the refracted pulse is subject to mode conversion at the vicinity of the null point \citep{McLaughlin2006b}, leading to the generation of slow-wave trains along the spines of the magnetic skeleton (i.e. the separatrices) \citep{Santamaria2018, Tarr2019ApJ...879..127T}. Alongside these slow waves, we also see velocity perturbations propagating across the magnetic field, a behaviour indicative of fast waves in low-$\beta$ plasma.

\begin{figure*}[t]
    \centering
    \includegraphics[trim={0.0cm 0.cm 0.cm 0.cm},clip,scale=0.55]{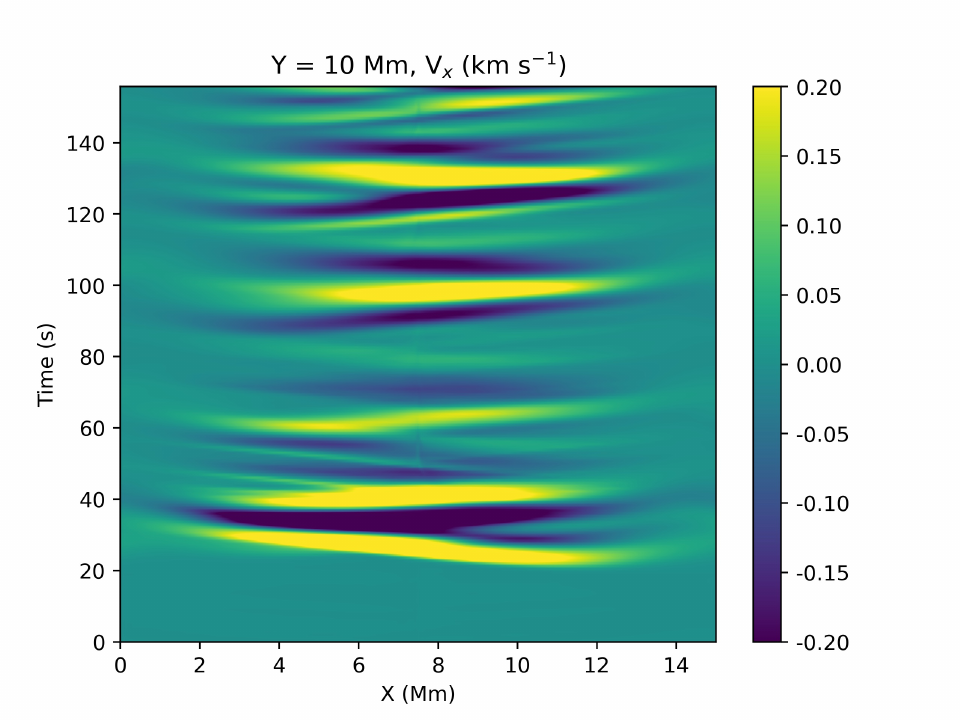}
    \includegraphics[trim={0.0cm 0.cm 0.cm 0.cm},clip,scale=0.55]{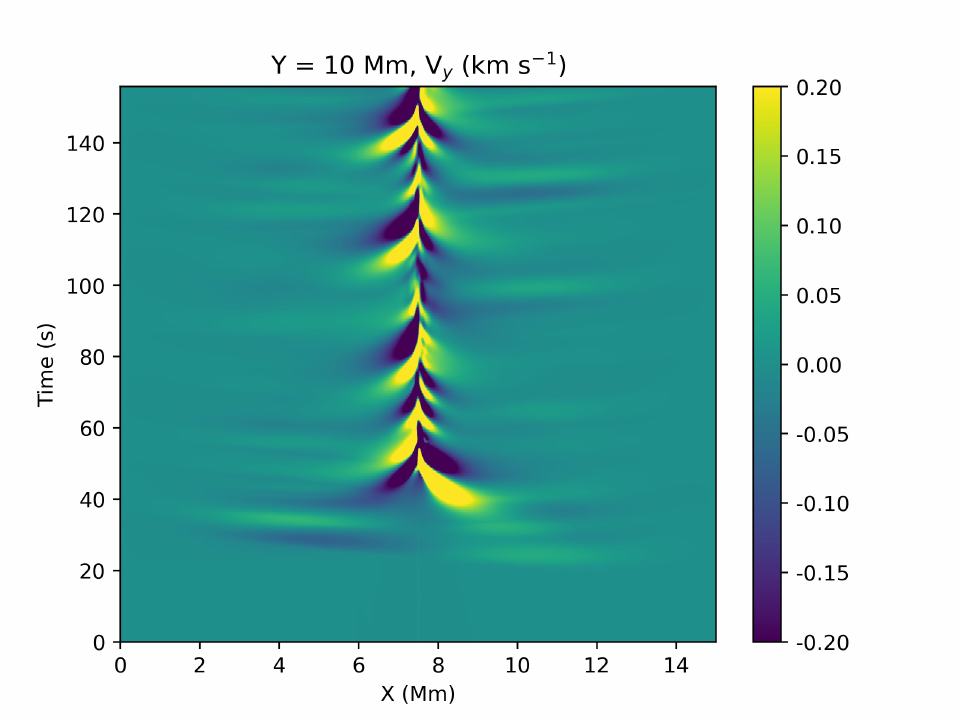}
    \caption{Time-distance profiles of the $V_x$ (left panel) and $V_y$ (right panel) velocity components, along a horizontal slit at $Y=10$ Mm, for model M1.}    \label{fig:xtslits}
\end{figure*}

In Figure \ref{fig:ztslits} we present the time-distance maps of a vertical cut along the $X=7.5 $ Mm line, corresponding to the separatrix our magnetic null point, for the $V_x$ and $V_y$ velocity components. Alongside them we show two snapshots of MHD wave proxies \citep{Raboonik2022thesis, Enerhaug2024A&A...681L..11E} for the compressible parallel component ($\mathcal{C}_{||}$, top right) and the compressible transverse component ($\mathcal{C}_{\perp}$, bottom right):
\begin{align}
    &\mathcal{C}_{||}= \nabla\cdot(v_{||}\mathbf{e}_{||})\\
    &\mathcal{C}_{\perp} = \nabla \cdot \mathbf{v} - \mathcal{C}_{||},
\end{align}
where $\mathbf{e}_{||}$ is the unit vector parallel to the magnetic field and $v_{||}$ is the velocity component parallel to the field. Although these wave
identifiers do not coincide with the definitions of the slow and fast MHD wave modes, they are associated with important distinguishing characteristics of these two modes. In low-$\beta$ plasma, the slow and fast modes corresponds well with $\mathcal{C}_{||}$ and $\mathcal{C}_{\perp}$, respectively. In high-$\beta$ plasma, however, features of the pressure waves can appear in $\mathcal{C}_{\perp}$, therefore additional information, such as the propagation velocity needs to be considered in the analysis. Following \citet{Enerhaug2024A&A...681L..11E}, we will use the terms slow ($\mathcal{C}_{||}$) and fast ($\mathcal{C}_{\perp}$) identifiers, for brevity.

The $V_x$ component, i.e. the one across the vertical slit, traces the behaviour of the wave pulse that perturbs the null point. As we see from the inclination of the velocity strips away from the equipartition layer (dotted white lines), the initial pulse can be mostly described as a fast wave that first reaches the vertical separatrix at about $Y=3.5$ Mm, at $\sim 18$ s, only to then converge to the null point. This agrees with the nature of the incoming, perturbing pulse as a fast wave, as shown by comparing the slow and fast wave identifiers at the $t=17.91$\,s in the accompanying panels. As the refraction around the null point takes place, we see a series of upward and downward propagating fast wave pulses, in the $\beta<1$ region. In the $\beta>1$ area, the inclinations of the waves change. This effect is expected to be caused (a) from the fast to slow wave mode conversion that takes place near the equipartition layer \citep{McLaughlin2006b} and (b) from the reconnection outflows, as they have been observed in \citet{McLaughlin2009}. The evidence for the reconnection will be discussed in the following figures. In the same panel we also plot the trajectory (red dashed-dotted lines) of a  wave propagating with the local Alfv\'{e}n speed (at $t=0$) along height. We plot two trajectories, one starting from $Y=0$ at $t=0$ and one from $Y=5.6$ Mm (right above the null point) at $t=0$. This is done for better visualization, due to the fact that the Alfv\'{e}n speed drops to zero at the null, and thus the trajectory there would practically become a horizontal line. We see that the inclination of the $V_x$ component is closely matching that of a wave travelling with the Alfv\'{e}n speed, implying that these $V_x$ strips along the $X=7.5$ Mm line are propagating faster than the slow waves.

In the same Figure, we see the fast wave (traced by $V_x$ to propagate through the transition region without any significant reflection. In \citet{NakariakovRoberts1995SoPh..159..399N} it was shown how propagating fast waves in an inhomogeneous coronal plasma are refracted into regions of low Alfv\'{e}n speed. \citet{McLaughlin2004} used the same principle to explain the refraction of coronal fast waves towards a null point. In the case of our model, even though $\beta<1$ in the chromospheric part, the Alfv\'{e}n speed drops more than an order of magnitude drop as we cross the transition region from the corona towards the chromosphere. The fast waves essentially experience refraction towards the chromosphere, effectively fully transmitting through the transition region. Once the fast waves cross the transition region, they reach the bottom boundary where they get reflected due to our choice of boundary conditions. We then observe a series of reflections taking place between the transition region and the bottom boundary of our domain, which in turn lead to additional pulses being generated and released in the corona. Unlike the downward  fast waves propagating downward from the corona, the upward propagating fast waves that hit the transition region at an oblique angle experience severe refraction and are turned back toward the lower atmosphere, with only a small fraction of waves, which propagate near vertically, making it to the corona. A similar effect of pulses being reflected at the lower boundary and returning to the corona has been observed in the past by \citet{Tarr2019ApJ...879..127T} in a similar setup, and leads to additional pulses reaching the null point, resulting in refraction, mode conversion and perturbing the magnetic field at the null point from its equilibrium state at periodic intervals. 

The bottom panel of Figure \ref{fig:ztslits} shows the $V_y$ component, which is along this vertical slit. Here we see a series of upward and downward propagating waves generating in the vicinity of the null point, resulting from the interaction of the initial velocity pulse with the magnetic null point. A similar behaviour is observed in a snapshot of the slow wave identifier at $t=155.71$\,s, which clearly shows the slow wave trains propagating away from the null point, along the magnetic field at the separatrices. In the same panel we also plot the trajectory (red dashed line) of a wave travelling with the local sound speed (at $t=0$) along height. The inclination of the $V_y$ component is almost parallel to that of the sound wave, implying that these $V_y$ strips along the $X=7.5$ Mm line are slow MHD waves. We also see here that for the slow waves that propagate downwards towards the chromosphere, once they reach the transition region boundary, the sudden jump to high density-low temperature plasma leads to their reflection back into the corona, with only traces of slow waves being detected to propagate downwards in to the chromosphere.

Studying the periodicity of these signals, we plot the time series for $V_x$ and $V_y$ at three different heights along the same vertical slit and we calculate the respective Fourier frequencies, as shown in Figure \ref{fig:velocityFourier}. We report a main frequency for the $V_x$ signals, centred around $57.5$ mHz, and another one for the $V_y$ signals around $63.4$ mHz. Although it is not clear why these two main frequency bands are not an exact match to each other, they are clearly consistent among the respective signals at different heights. In addition, both of them are close the frequencies of the wave trains reported in \citet{Santamaria2017} for a similar setup of a coronal null point, implying that they are probably related to the background plasma conditions, as we will further explore later in the manuscript. In addition, we also observe a higher frequency band near $90$ mHz, as well as a very strong signal at $\sim 25.6$ mHz for the $V_x$ component at $Y=3$ Mm, that sits well above the $95\%$ confidence level. The $V_y$ component also shows a faint peak at $\sim 31$ mHz, which becomes weaker at higher heights. The latter frequency band, which is a lot weaker at higher heights, is associated with the additional pulses generated by the wave reflections at the bottom boundary. In the right panels of Figure \ref{fig:velocityFourier} we plot the wavelet spectra over time for the $V_x$ and $V_y$ components at the $Y=3$ Mm, along the same vertical slit ($X=7.5$ Mm). The $V_x$ component shows a strong persistent signal near $25$ mHz, another strong persistent signal at $\sim 56$ mHz and a high-frequency, noise-like signal at $\sim 90$ mHz. On the other hand, $V_y$ clearly shown one main frequency band between $60$ and $70$ mHz, as well as a very faint frequency band at $\sim 30$ mHz, similarly to the Fourier spectra shown in the same Figure. These frequencies extend to the entirety of the signal, before and after the slow waves reflected from the transition region reach again the null point, at $\sim 70$ s. Finally, in the top left and middle panels, we also include the $V_x$ component at $X=10$ Mm and $Y=5.05$ Mm (red line). This point is on the ``horizontal’’ separatrix, to the right of the null point. At that location, the $V_x$ velocity traces waves travelling along the field line. The frequency of these waves are matching those generated at the null point, similarly to the $V_y$ signals at larger heights matching the one at the null. We see a clear periodicity imposed by the resonant cavity at the null point, on top of any driving imposed by the reflected waves propagating upwards from the bottom boundary and/or transition region, hinting at a connection between the detected waves away from the null point, and the resonant cavity in our system, as reported in \citet{Santamaria2018}.

\begin{figure*}[t]
    \centering
    \includegraphics[trim={0.cm 0.cm 0.cm 0.cm},clip,scale=0.55]{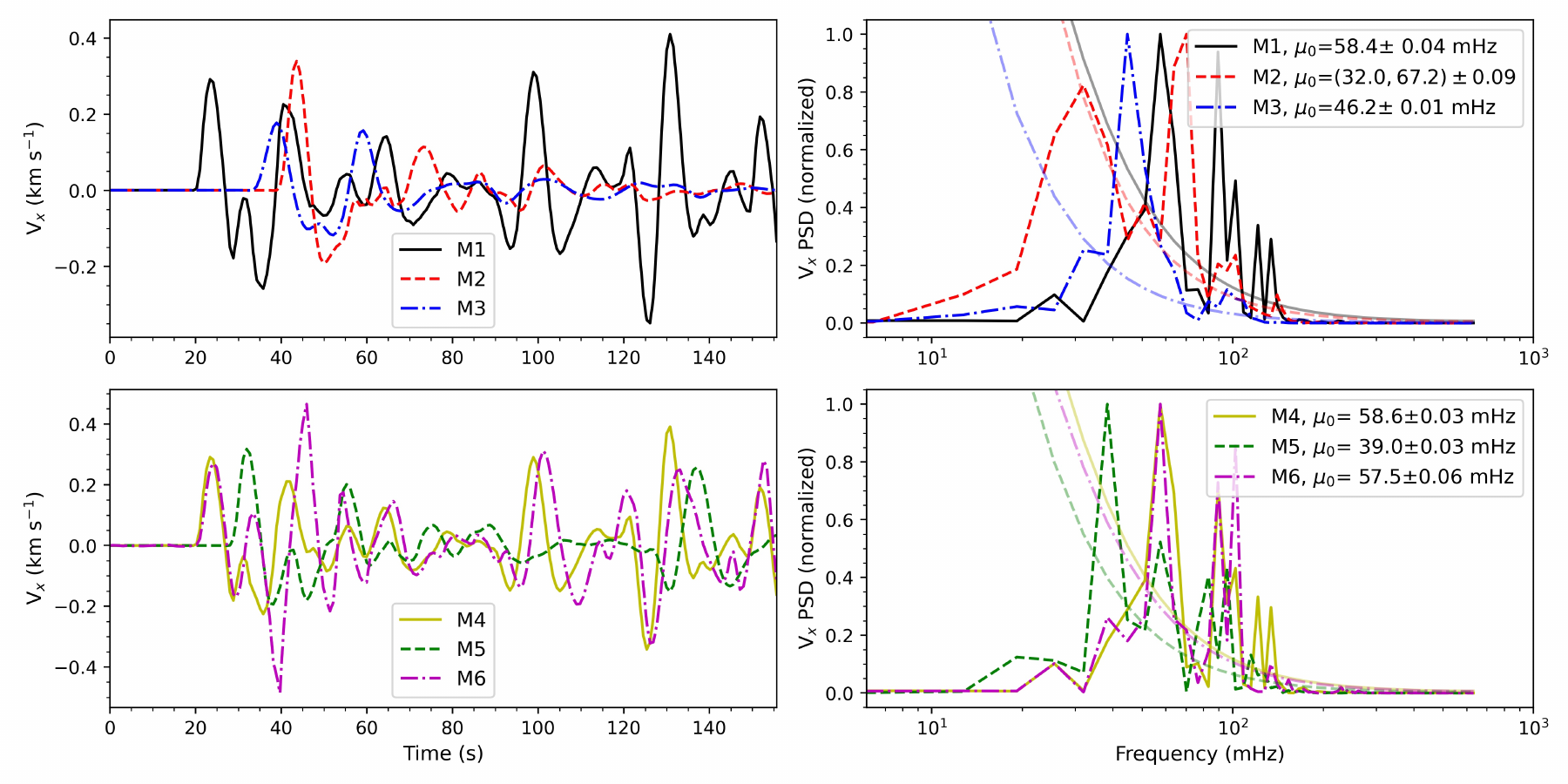}
    \caption{Time series of the $V_x$ velocity components at the $(X,Y)=(10,10)$ Mm point for all six of our models and their respective Fourier spectra, alongside the $95\%$ confidence level curves corresponding to each spectrum, in matching, less opaque lines.}    \label{fig:parameterQFPs}
\end{figure*}

From \citet{McLaughlin2009} we expect a null point perturbed by a magnetic pulse to be undergoing oscillatory reconnection. Our setup does not have the necessary resolution to properly resolve the generated current sheets and the reconnection  dynamics. However, one way to study the manifestation of reconnection in our setup is to track the evolution of the magnetic field lines at the null point, tracing them from the field footprints, in the bottom boundary. In figure \ref{fig:nullpointfieldlines} and the accompanying animation, we show the magnetic field line evolution over time, alongside the $J_z$ current density. The white and black lines represent the field lines in the opposite ``quadrants’’ around the null, defined by the separatrixes. For a black (white) field line to enter the adjacent quadrant, the field topology needs to break through reconnection \citep[see also][]{McLaughlin2009}. Our setup shows evidence of periodic reconnection, although its nature as driven reconnection as opposed to oscillatory reconnection needs to be understood by studying the frequencies associated with the reconnection. 

To trace the reconnection signatures, in Figure \ref{fig:nullpointreconnection} we plot the signals of the average $J_z$ current density at the null point \citep{Karampelas2022a,Karampelas2022b,Karampelas2023ApJ...943..131K}: 
\begin{align}
    &\langle J_z \rangle = \frac{1}{A} \int_A J_z(t) \, dA, \, x\in[7.4,7.6],\, y\in[5.45,5.65]\text{ Mm},
\end{align}
with $A$ being the surface area of the square over which we integrate, as well as the magnetic flux rates for the $B_x$ and $B_y$ components across lines at $X = 7.4$ Mm ($\dot{\Phi}_B(X)$) and at $\Delta Y = 0.1$ Mm above each null point ($\dot{\Phi}_B(Y)$) respectively:
\begin{align}
    &\dot{\Phi}_B(X) = \frac{d}{dt} \int_{Y_{null}-0.1\, Mm}^{Y_{null}+0.1\, Mm} B_x(X=7.4\, Mm, Y, t) \, dY \\
    &\dot{\Phi}_B(Y) = \frac{d}{dt} \int_{X_1=7.4\,Mm}^{X_1=7.6\, Mm} B_x(X, Y=Y_{null}+0.1\, Mm, t) \, dX.
\end{align}
We note here that initial, equilibrium location of each null point along the Y-axis depends on the model (M1 to M6). The magnetic flux rates, being the equivalent of the reconnection rate \citep{McLaughlin2009,Tarr2019ApJ...879..127T}, trace the evolution of the magnetic field as the latter is ``pushed’’ by the lateral movement of the separatrices due to the periodic reconnection. From the Fourier and wavelet spectra, we see that the current density density shows a strong persistent signal matching the low frequency driving signal from the reflected fast waves ($\sim 25$ mHz), as well as a weaker persistent signal ($\sim 56$ mHz), at frequencies matching the $V_x$ component at $(X,Y)=(7.5, 3)$ Mm, studied in Figure \ref{fig:velocityFourier}. On the other hand, the $\dot{\Phi}_B(X)$ and $\dot{\Phi}_B(Y)$ flux rates show strong persistent signals at around $56.6$ mHz and $63.7$ mHz respectively, and weaker persistent frequencies at around $25.5$ and $30$ mHz respectively, matching the frequencies of the $V_x$ and $V_y$ velocity components at the null point. Our results indicate that the periodicity of the reconnection  closely follows that of the waves generated at the null point, with a smaller contribution being attributed to the driving of the external fast wave perturbation. Finally, for all three of the signals, higher frequency noise is also generated at different parts of the time series.

\subsection{Dependence of periodicity to plasma configurations}
Expanding our analysis to the other models M2 to M6, we first show the 2D profiles of the $V_x$, $V_y$ and $V_z$ velocity components, at $t=155.71$\,s for the six models M1 to M6 in Figure \ref{fig:comparative}. Just like in Figure \ref{fig:alternative}, the magnetic field lines, top of the transition region and the $\beta=1$ layer around the each null point are also included in the panels. Similarly to Figure \ref{fig:2Dvelocities} for M1, we see that for all six models the refracted pulse is subject to mode conversion at the vicinity of each null point. This leads to the generation of slow-wave trains along the spines of the magnetic skeleton (i.e. the separatrices) to velocity perturbations propagating across the magnetic field, a behaviour indicative of fast waves in low-$\beta$ plasma. In the case of M4, we also see an additional non-zero $V_z$ component. As described earlier, M4 is the only setup where we have included a $B_z$ component of the magnetic field, i.e. a component in the ignorable direction, making this a 2.5D model, rather than a purely 2D model. This $V_z$ velocity component is the result of mode conversion of the slow and fast magnetoacoustic perturbations to Alfv\'{e}n waves due to the background guide field \citep{Bulanov1992PPCF...34...33B,McClements2006JPlPh..72..571M,Landi2005ApJ...624..392L}, and it propagates along the field lines, asymptotically converging to the separatrices \citep{BulanovSyrovatskii1980FizPl...6.1205B, HassamLambert1996ApJ...472..832H}.

We compare the Fourier spectra of the $\dot{\Phi}_B(X,Y)$ reconnection rates with the spectra of the $V_{x,y}$ velocities at each null point, as shown in Figure \ref{fig:parametercavity}. Just as in the case of the M1 model (shown in the top left of the Figure), we observe a clear connection between the derived frequencies of the reconnection and of those of the planar velocity components generated at the null point. This further supports the notion that the resonant cavities, as defined by the coronal null points, are connected to the oscillatory reconnection properties of said magnetic null points. In other words, the wave dynamics from these resonant cavities, are intrinsically connected to the relaxation mechanism of oscillatory reconnection. This supports our hypothesis that these two phenomena are interconnected, with the null point imposing its properties both on the reconnection and the generation of waves in the resonant cavity.

Looking at model M4 in the bottom left panels of Figure \ref{fig:parametercavity}, we see that the third velocity component ($V_z$, i.e. the Alfv\'{e}n wave) away from the null ($Y=3$ Mm and $Y=8$ Mm) shows similar frequencies to the fast and slow waves (as traced here by $V_x$ and $V_y$, respectively) associated with the reconnection process at the null ($\dot{\Phi}_B(X)$ and $\dot{\Phi}_B(Y)$, respectively). As stated earlier, the third velocity component is the result of mode conversion of magnetoacoustic perturbations to Alfv\'{e}n waves due to the background guide field. This explains why the Alfv\'{e}n wave along the separatrix has frequencies similar to the fast and slow waves associated with the resonant cavity and by extention with the reconnection process. Looking at the Alfv\'{e}n wave signal at the null point ($X=7.5$ Mm and $Y=5.5$ Mm), we see two prominent peaks. The one at $\sim 30$ mHz, although falling below the $95\%$ confidence level, is reminiscent of the low frequency fast wave that propagates to the null from the chromosphere, as shown in Figures \ref{fig:ztslits} and \ref{fig:velocityFourier}. The second one, at $\sim 90$ mHz follows the high frequency peak of $\dot{\Phi}_B(X)$. However, there is no clear correlation between the frequencies of Alfv\'{e}n wave signal at the null (blue, dashed-dotted line) and the main frequency peaks of $\dot{\Phi}_B(X)$ and $\dot{\Phi}_B(Y)$ at $\sim 57$ and $\sim 63$ mHz, respectively. As the Alfv\'{e}n wave propagates along the magnetic field lines, it cannot cross the null point even with a $B_z$ field component present, since the topology of planar field lines will break at the null point. What we measure at the fixed point $(X,\,Y) = (7.5,\, 5.5)$ Mm is the combined signal of wave mode conversion and the Alfv\'{e}n waves propagating along the field lines when the magnetic null point is perturbed away from that fixed point. This seems to lead to a complicated signal that prevents a clear correlation between the frequencies of the Alfv\'{e}n wave and the reconnection process, when measuring directly at the null.

\subsection{Propagating waves generated by the reconnecting null point}
As a final point, we want to further emphasize the connection between the wave generation from the reconnecting null point and the propagating fast waves further away from these resonant cavities. In Figure \ref{fig:xtslits}, the $V_x$ and $V_y$ velocity components across the domain over time, at $Y = 10$ Mm are shown. At that height, the magnetic field is almost vertical, as seen from Figures \ref{fig:inicon} and \ref{fig:alternative}, for all models. Therefore, the $V_x$ component will mostly track the waves propagating perpendicular to the field, i.e. the fast waves. This is can also be seen from the snapshots of the fast and slow wave identifiers in Figure \ref{fig:ztslits}, where we can see the fast waves propagating across the domain while the slow waves are mostly concentrated along the magnetic field spine of the null point. In the left panel of Figure \ref{fig:xtslits} we see multiple wave fronts propagating across the domain, as a result of the refracting fast waves seen in Figures \ref{fig:2Dvelocities} and \ref{fig:ztslits}. Similar, fainter signatures are shown in the right panel of Figure \ref{fig:xtslits}, for the $V_y$ velocity component, although the main signal there is attributed to the slow wave train propagating along the $X=7.5$ Mm separatrix. Picking a random point $(X, Y)=(10, 10)$ Mm away from the $X=7.5$ Mm separatrix, we find the frequencies of these propagating fast waves from the $V_x$ velocity signals, for all six of our models. From the velocity signals and Fourier spectra in Figure \ref{fig:parameterQFPs}, we again see a close relation between the frequency of the fast waves and the velocity signals generated at each null point. Therefore, these propagating waves appear to have the same periodicity to the one imposed by the resonant cavity at the perturbed, reconnecting null points.


\section{Discussion and Conclusions} \label{sec:discussions}
It has been well established that coronal null points act as resonant cavities, generating waves the frequencies of which depend on the background plasma conditions, when perturbed by an external wave \citep{Santamaria2016,Santamaria2017,Santamaria2018}. These generated waves are understood as the result of mode conversion of incoming waves in the vicinity of null points, as has been shown in multiple null point configurations in the linear regime \citep{McLaughlin2011SSRv}. At the same time, reconnecting null points without an external driver are characterised by intrinsic properties regarding the periodicity of the excited reconnection process \citep{Karampelas2022b,Karampelas2023ApJ...943..131K}, which in turn creates Alfv\'{e}n, fast magnetoaccoustic and slow magnetoaccoustic waves. Both of these mechanisms have the potential to be further employed in seismological studies in the solar atmosphere. By perturbing the null points across all our six setups with a single velocity pulse (plus the reflected pulses, see below), our systems generate a series of higher frequency propagating slow and fast waves, through a combination of wave refraction and mode conversion. The frequencies of the resulting waves differ per case, depending upon the background plasma conditions per case, similarly to the results of \citet{Santamaria2018}. The frequencies of the generated waves also match those of the resulting periodic reconnection, as measured by the magnetic flux rates, in the vicinity of each null point. Our results supports our initial hypothesis that the two phenomena, i.e. the resonant cavity and the periodic reconnection at null points are interconnected.  Past studies have considered the periodicity imposed by the wave-generating resonant cavity to match that of oscillatory reconnection \citep{Schiavo2024ApJ...975...10S}. However, this is the first study producing evidence of this matching periodicity between oscillatory reconnection and the wave-generating resonant cavity related to the null point. A summary of these results can be found in Table \ref{table:2}, for each model.

\begin{table}
\caption{List of setups and corresponding frequencies.}              
\label{table:2}      
\centering                                      
\begin{tabular}{l c c c c}          
\hline\hline                        
Model & $\dot{\Phi}_B(X)\,f_0$ & $\dot{\Phi}_B(Y)\,f_0$ & $V_x\, f_0$ & $V_x\, f_0$  \\    
\hline                                   
    M1 & 56.9 mHz & 63.8 mHz & 57.5 mHz & 64.1 mHz \\      
    M2 & 33.2 mHz & 32.3 mHz & 32.3 mHz & 31.6 mHz \\
    M3 & 45.0 mHz & 50.0 mHz & 45.3 mHz & 50.0 mHz \\
    M4 & 56.9 mHz & 63.3 mHz & 57.5 mHz & 63.5 mHz \\
    M5 & 38.4 mHz& 45.1 mHz & 38.4 mHz & 45.2 mHz \\
    M6 & 55.8 mHz & 61.8 mHz & 57.0 mHz & 61.3 mHz \\
\hline                                             
\end{tabular}
\tablefoot{Main frequencies ($f_0$) for the spectra of the magnetic flux rates $\dot{\Phi}_{x,y}$ and of the $V_{x,y}$ velocity components at the null point of each model M1 to M6.}
\end{table}

Past studies of periodically reconnecting coronal null points \citep{Heggland2009ApJ,Tarr2017ApJ...837...94T,Tarr2019ApJ...879..127T} have focused on the dynamics, energy evolution and mode conversion in cases of forced reconnection. However, none of these studies had focused on the manifestation of oscillatory reconnection as a relaxation process nor had detected signatures of the resonant cavity in the reconnecting magnetic null points. This present study not only shows the connection between periodic reconnection of null points and the resonant response of null points to external waves, but we have also seen how the null points impose their own periodicity over that of an external lower-frequency driving, such as the reflected fast waves from the bottom boundary seen in our simulations. In particular, these reflected waves are essentially an external driver, the frequency of which ($\sim 26$ mHz) can be detected in the wavelet spectra of incoming fast waves near the null point, with $95\%$ confidence and was shown to be contributing to the time series for the entirety of the simulation. However, the same frequency is only detected with significant power in the average out-of-plane current density, which has been used in the past as a tracer for reconnection. Looking at the evolution of the magnetic flux rate across slits near the null points, the same frequency is very weak in the signals, falling outside the $95\%$ confidence interval. On the contrary, the main frequencies that are detected from the magnetic flux rates, as well as the average $J_z$ current density, match well with the main frequencies of the generated slow and fast waves. By having the null point imposing its intrinsic properties both on the reconnection process and on the wave generation process, we are led to characterise this periodic reconnection as oscillatory reconnection and stress the latter's connection to the resonant cavity.

We need to note here that our setups do not properly resolve the reconnection current sheets of the collapsed null points, nor do they include explicit physical resistivity, only effective numerical dissipation. Thus, our setups offer little insight on the finer dynamics of the periodic reconnection and how they affect, or are affected by the response of the resonant cavity to external excitation. However, it has been shown that the levels of resistivity do not affect the period of oscillatory reconnection, only the maximum amplitude of the current density, the nature of its decay rate and the magnitude of ohmic heating at the null point \citep{Talbot2024ApJ...965..133T}. Therefore, including resistivity in our setups would not change the reconnection frequencies found here. 

Finally, the velocity snapshots of Figure \ref{fig:2Dvelocities}, the time-distance velocity profiles of Figure \ref{fig:xtslits} and the velocity time series and Fourier spectra of Figure \ref{fig:parameterQFPs} imply a connection between the fast waves generated by our resonant cavity - reconnecting null point systems and the quasi-periodic fast-propagating (QFP) magnetosonic waves found in observations. In \citet{LiuQFPW2011ApJ...736L..13L}, QFP wave trains imaged in EUV by SDO/AIA were detected in a flare and coronal mass ejection event. These waves were propagating outward along a funnel of coronal loops, with frequencies coinciding with those of the quasi-periodic pulsations of the flare, which in turn suggests a common origin between the two phenomena. Similar QFP waves were detected in a region of coronal condensations caused by magnetic reconnection between open and closed coronal loops \citep{LiLepingQFPW2018ApJ...868L..33L}. In that event, there was no observation of a flare, with the authors suggesting that periodic or oscillatory reconnection was responsible for these waves. Oscillatory reconnection has also been considered as a driver of QFP waves from an erupting flux rope \citep{2018ApJ...853....1S}. In a related study, \citet{KumarQFPW2017ApJ...844..149K} have reported quasi-periodic bursts in radio, microwave and soft X-ray emission associated with observed QFP waves. Both the quasi-periodic bursts and the QFP waves exhibited similar instant periods and were generated from quasi-periodic magnetic reconnection in the cusp region above flaring loops. Exploring this connection between QFP waves and periodically reconnecting null points, our results show velocity perturbations in the form of wave trains, propagating away from the null points both along and across the field. This is seen in Figures \ref{fig:2Dvelocities} and \ref{fig:xtslits}. In Figure \ref{fig:xtslits}, the wavefronts ($V_x$) across the domain exhibit slopes with respect to the time axis. This is due to propagation across the domain of the (fast) waves refracted around the null point \citep[e.g.][]{McLaughlin2006b}. These waves exhibit the same frequencies to those of the planar velocity components found at the null points in our different setups, from M1 to M6. Focusing on model M1, it is also apparent that the frequency of these waves does not much that of the low frequency driving by the fast waves reflected at the bottom boundary. As such, it is evident that these propagating fast waves in our models are generated by the response of the resonant cavity to external excitation and by extension, to the oscillatory reconnection process identified at the null points. These propagating fast waves seem to have the same driving mechanism and properties to the reported QFP waves in EUV observations, suggesting that they are in fact the physical mechanism behind these QFP waves. This also brings forth the possibility of using such QFP waves as tools for seismology. Their periods correspond to those of the periodic reconnection of null points that generated them and null points can impose a periodicity on the reconnection process. As such, QFP waves can be used to gauge the inherent periodicity of the periodically reconnecting null point system, which in turn can be used as diagnostic tools for solar coronal plasma, as was shown in \citet{Karampelas2023ApJ...943..131K}.

\begin{acknowledgements}
K.K. was supported by an FWO (Fonds voor Wetenschappelijk Onderzoek – Vlaanderen) postdoctoral fellowship (1273221N). K.K. also acknowledges the support by the DynaSun project under the UK Research and Innovation under the UK government the Horizon Europe funding guarantee EP/Y037456/1, and the ERC grant 101201424 (ACDCSUN). TVD received financial support from the Flemish Government under the long-term structural Methusalem funding program, project SOUL: Stellar evolution in full glory, grant METH/24/012 at KU Leuven. The research that led to these results was subsidised by the Belgian Federal Science Policy Office through the contract B2/223/P1/CLOSE-UP. It is also part of the DynaSun project and has thus received funding under the Horizon Europe programme of the European Union under grant agreement (no. 101131534). Views and opinions expressed are however those of the author(s) only and do not necessarily reflect those of the European Union and therefore the European Union cannot be held responsible for them. The computational resources and services used in this work were provided by the VSC (Flemish Supercomputer Center), funded by the Research Foundation Flanders (FWO) and the Flemish Government – department EWI. 
\end{acknowledgements}

\bibliographystyle{aa}
\bibliography{paper}
\begin{appendix}
\begin{onecolumn}
\section{High resolution run for M1 model} \label{sec:appendix}

\begin{figure}[h]
    \centering
    \resizebox{\hsize}{!}{
    \includegraphics[trim={0.cm 1.cm 3.6cm 0.5cm},clip,scale=0.55]{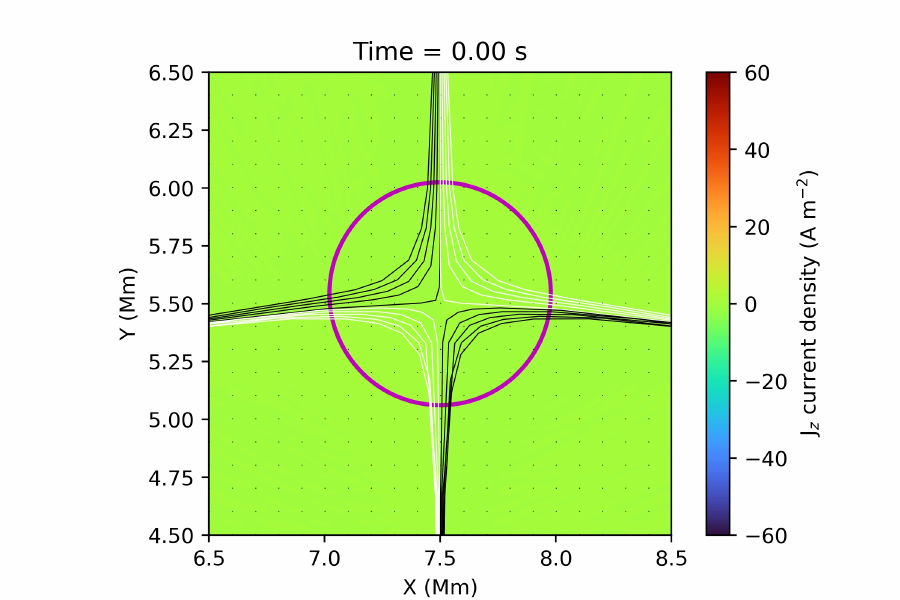}
    \includegraphics[trim={2.5cm 1.cm 3.6cm 0.5cm},clip,scale=0.55]{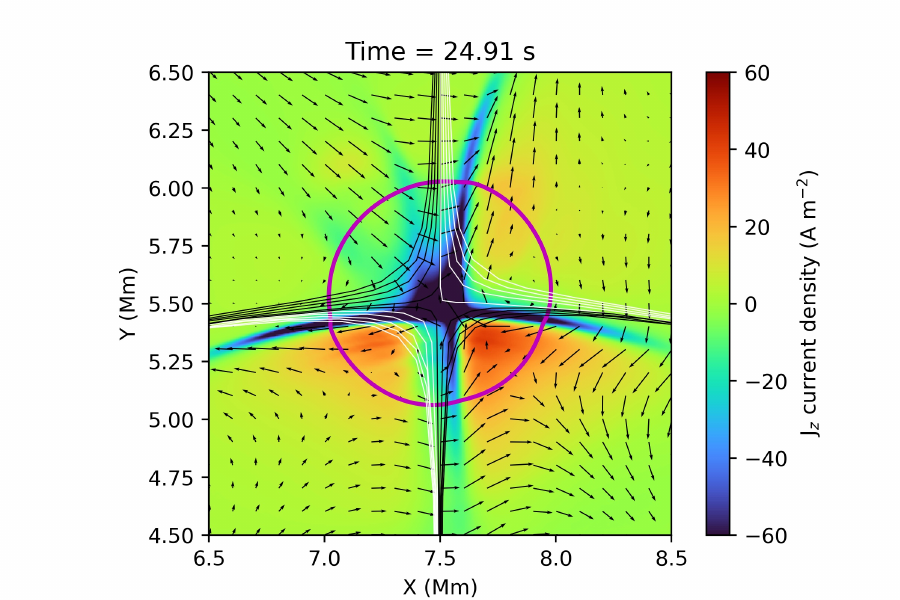}
    \includegraphics[trim={2.5cm 1.cm 3.6cm 0.5cm},clip,scale=0.55]{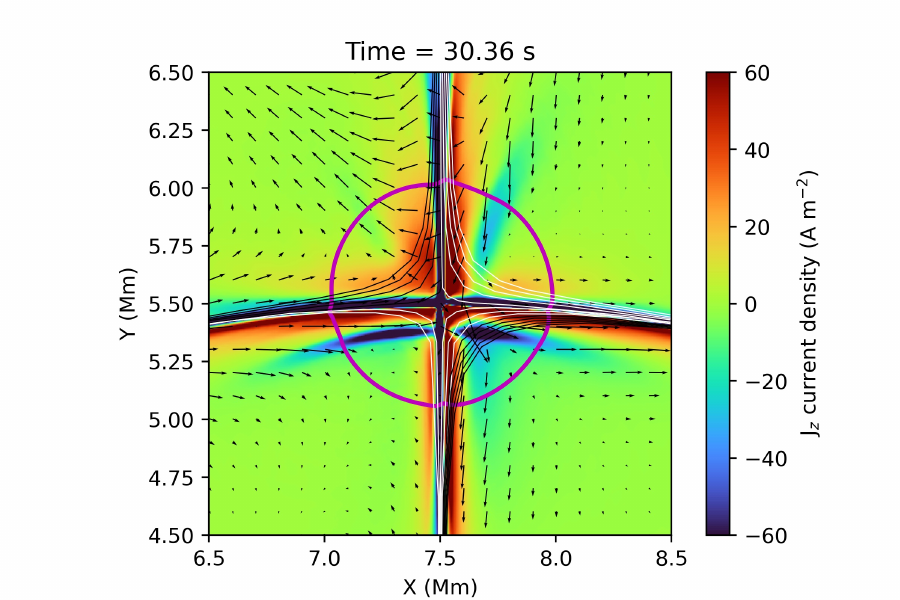}
    \includegraphics[trim={2.5cm 1.cm 3.6cm 0.5cm},clip,scale=0.55]{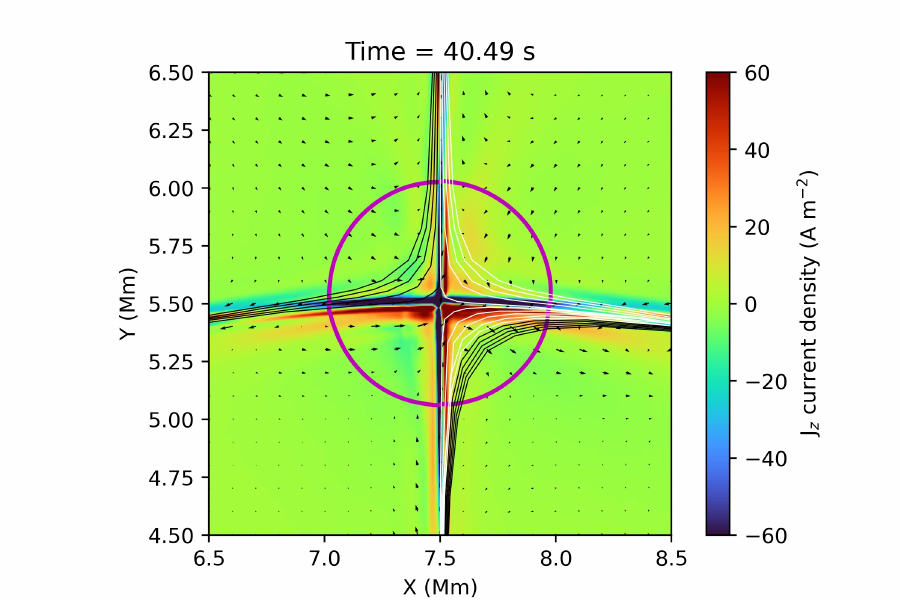}
    \includegraphics[trim={2.5cm 1.cm 0.cm 0.5cm},clip,scale=0.55]{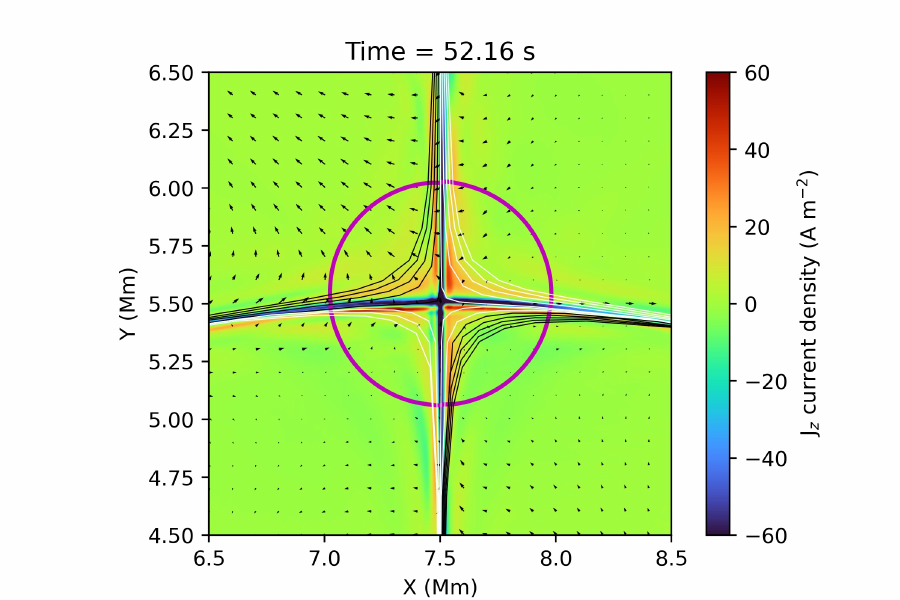}
    }
    \resizebox{\hsize}{!}{
    \includegraphics[trim={0.cm 0.cm 3.6cm 0.5cm},clip,scale=0.55]{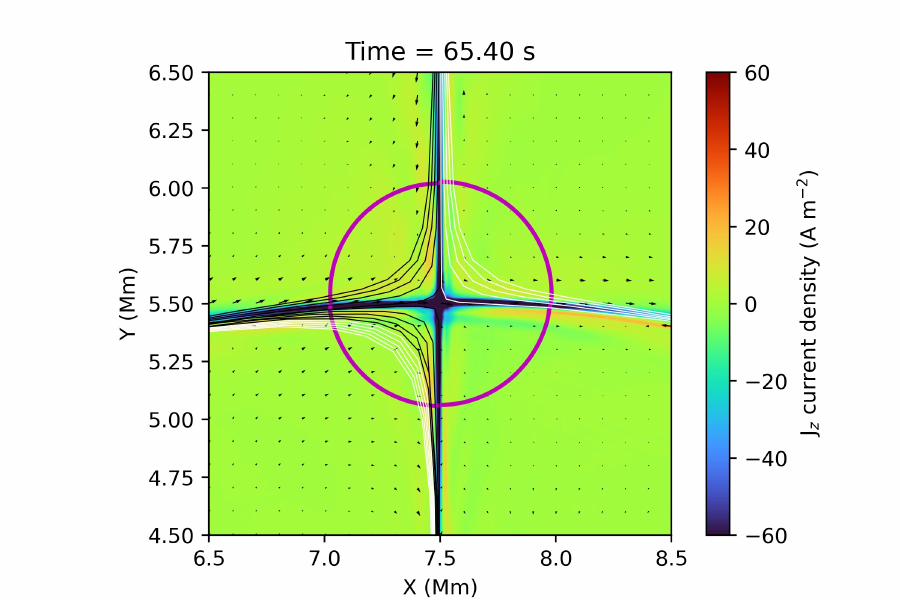}
    \includegraphics[trim={2.5cm 0.cm 3.6cm 0.5cm},clip,scale=0.55]{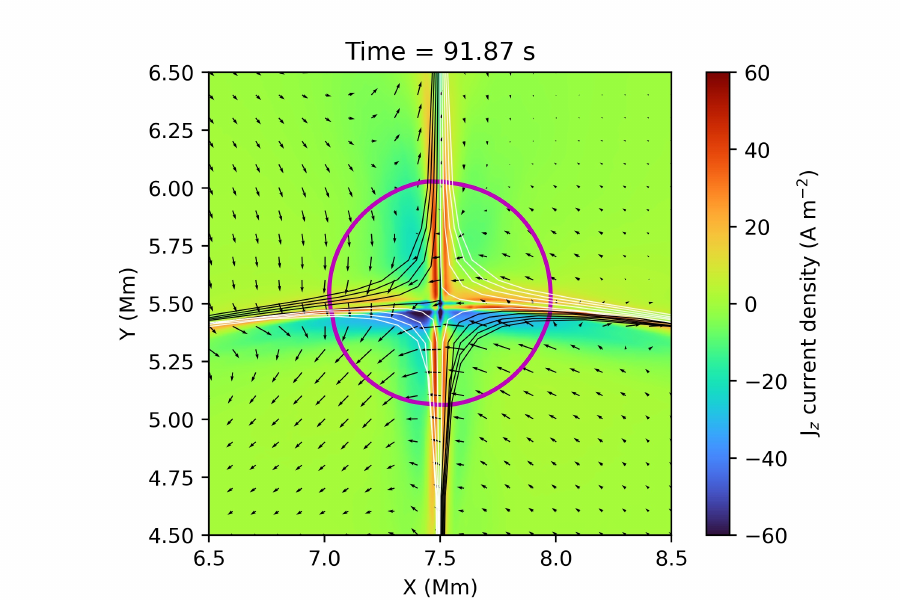}
    \includegraphics[trim={2.5cm 0.cm 3.6cm 0.5cm},clip,scale=0.55]{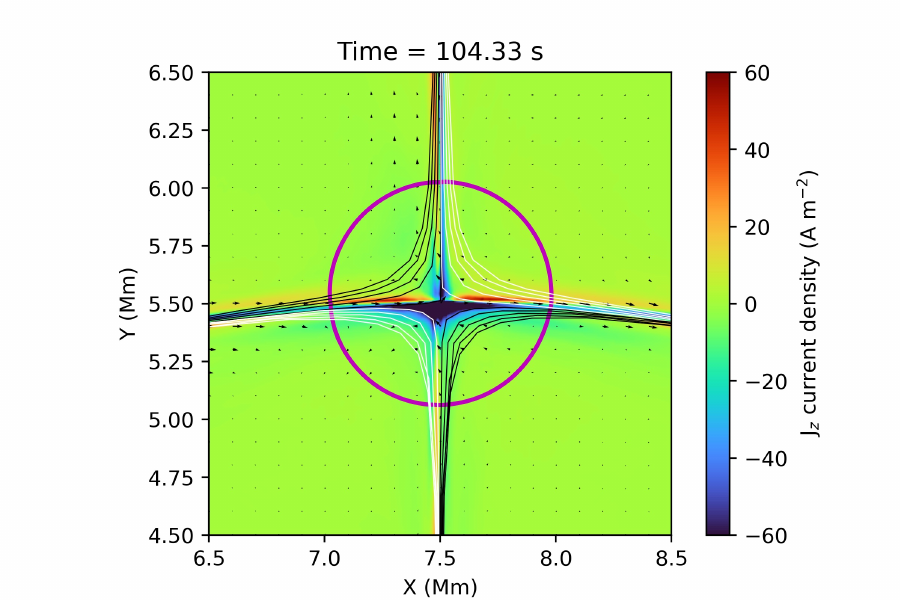}
    \includegraphics[trim={2.5cm 0.cm 3.6cm 0.5cm},clip,scale=0.55]{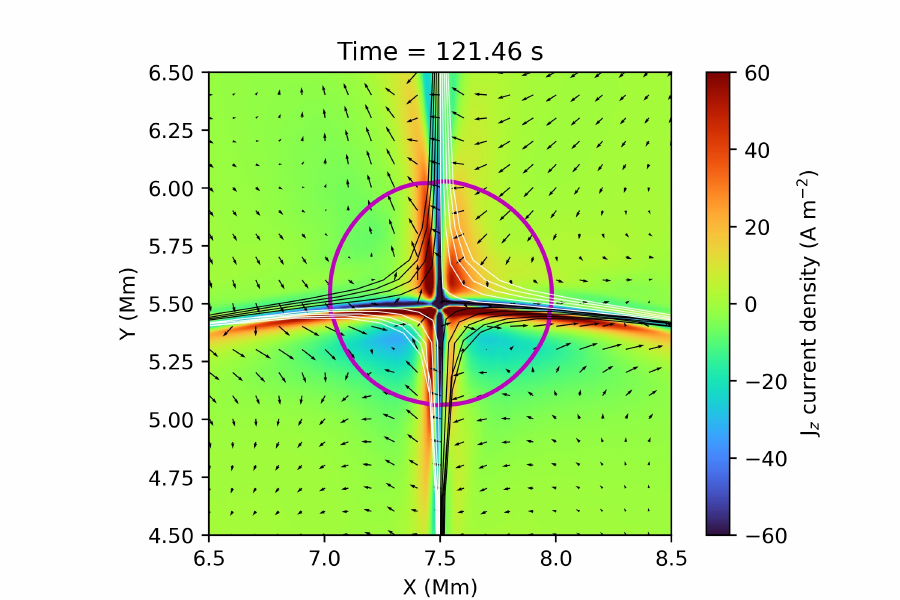}
    \includegraphics[trim={2.5cm 0.cm 0.cm 0.5cm},clip,scale=0.55]{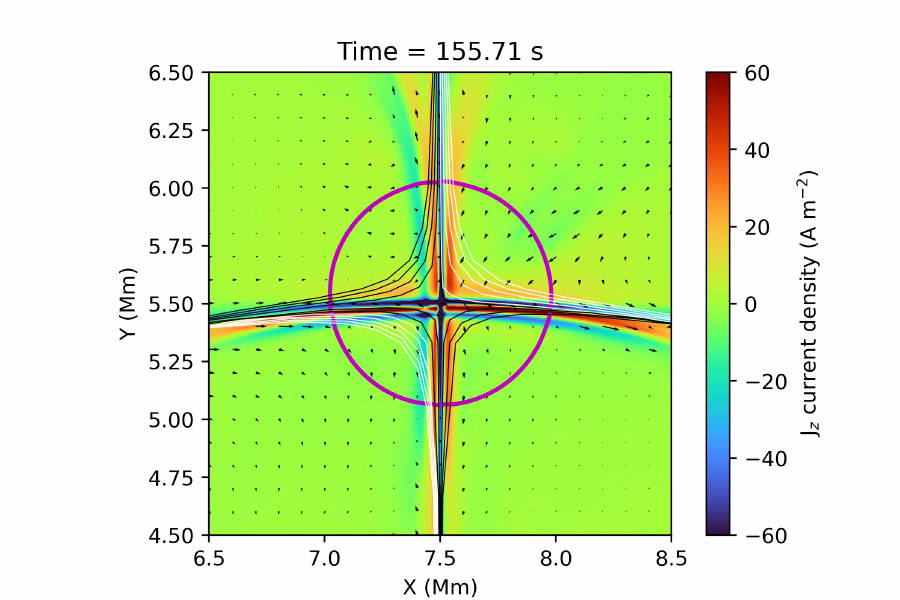}
    }
    \caption{Close-up snapshots of the $J_z$ current density at the null point during the reconnection events. Also shown here are the $\beta=1$ layer (purple circle), a sample of the magnetic field lines (in black and white) the vectors for the velocity field.}    \label{fig:nullpointfieldlines_hr}
\end{figure}

\begin{figure*}
    \centering
    \resizebox{\hsize}{!}{
    \includegraphics[trim={0.5cm 0.cm 1.cm 0.cm},clip,scale=0.55]{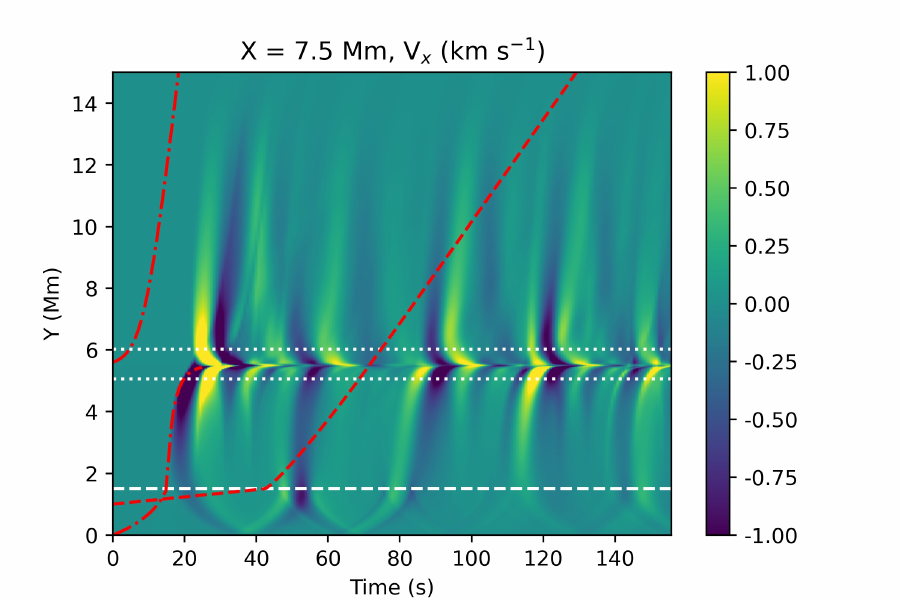}
    \includegraphics[trim={0.5cm 0.cm 1.cm 0.cm},clip,scale=0.55]{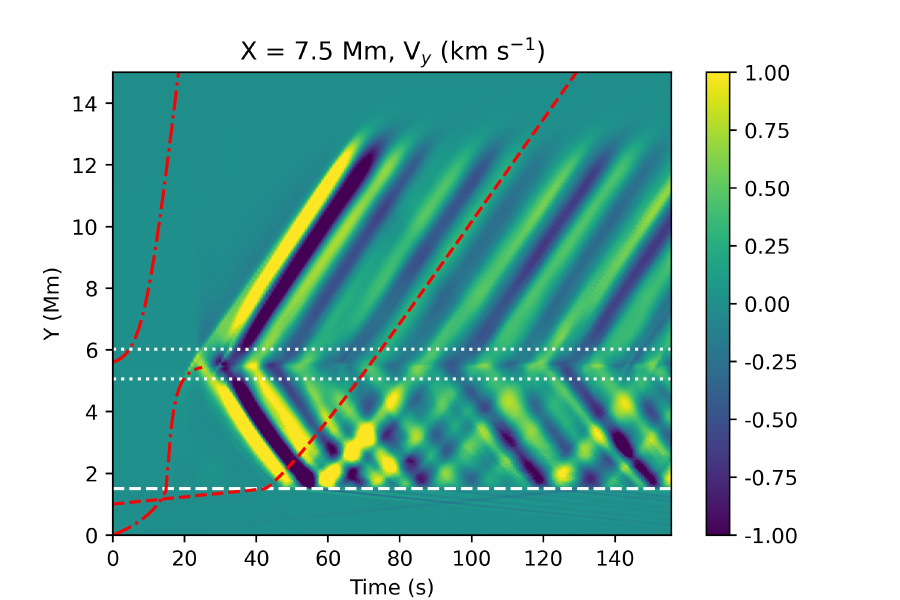}
    }
    \caption{Time-distance profiles of the $V_x$ (left panel) and $V_y$ (right panel) velocity components, similar to Figure \ref{fig:ztslits}, for the M1$_\mathrm{hr}$ model.} \label{fig:ztslits_hr}
\end{figure*}

\begin{figure*}
    \centering
    \includegraphics[trim={0.cm 0.cm 0.cm 0.cm},clip,scale=0.55]{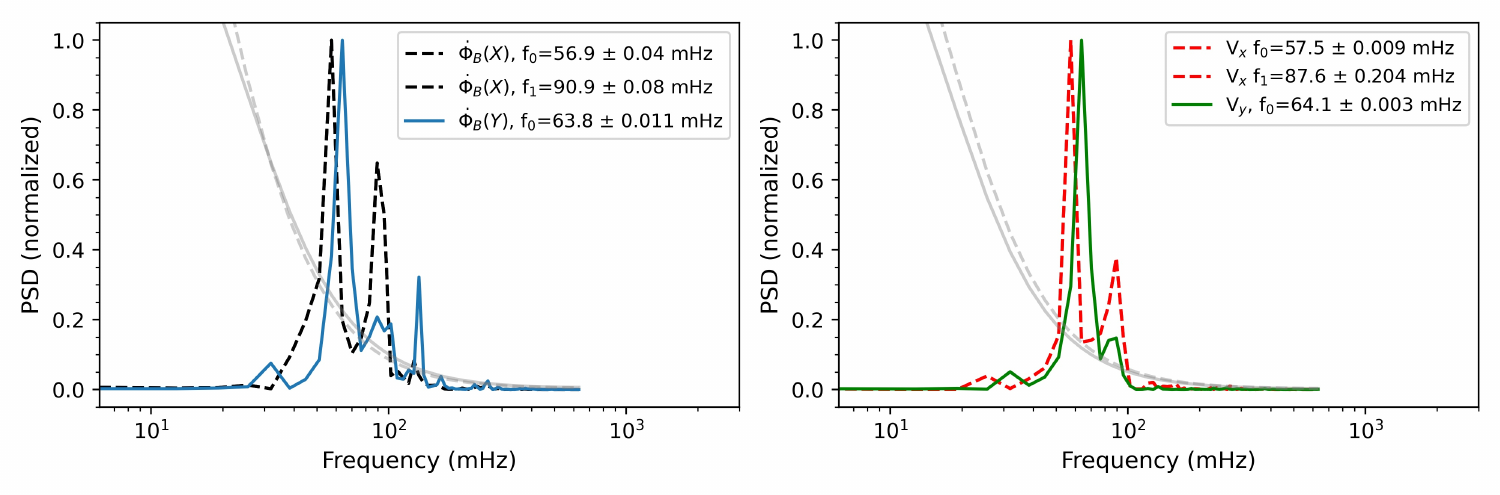}
    \\
    \includegraphics[trim={0.cm 0.cm 0.0cm 0.cm},clip,scale=0.55]{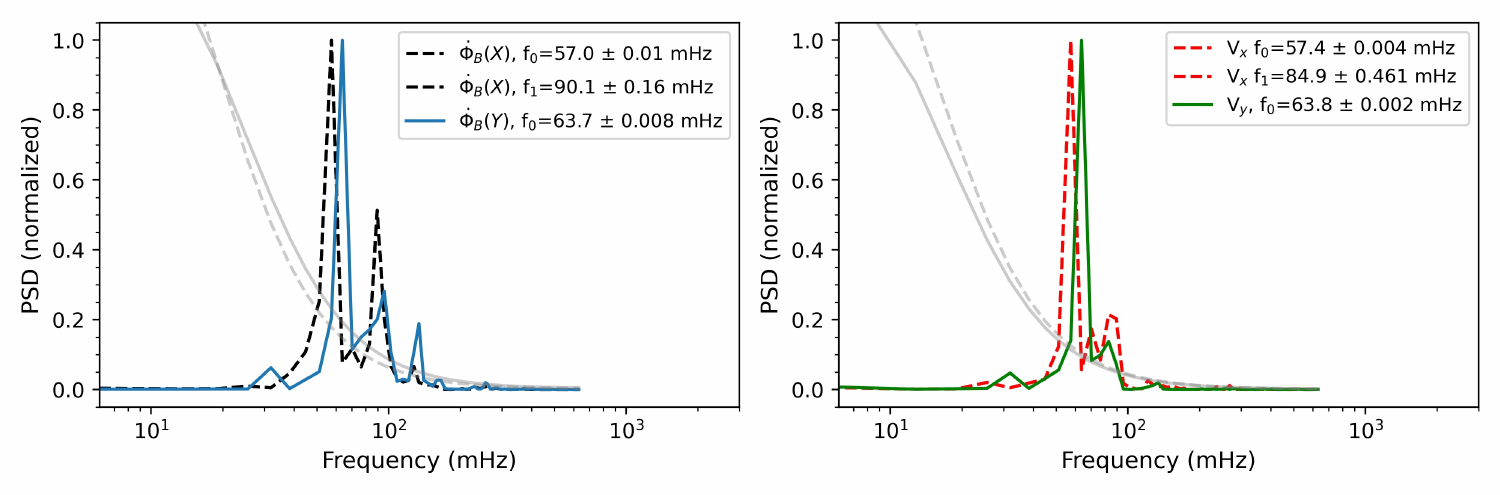}
    \caption{Fourier spectra of the magnetic flux rates $\dot{\Phi}_B(X,Y)$ and of the corresponding $V_{x,y}$ velocity components at the location of the null points, for models M1 (top panels) and for M1$_\mathrm{hr}$ (bottom panels). The $95\%$ confidence level curves are shown for each quantity in less opaque, matching style grey lines.}\label{fig:parametercavity_hr}
\end{figure*}

As stated in the main body of the text, the models that we have used do not have explicit magnetic diffusivity, but instead rely on the presence of effective numerical diffusivity, inherent to our numerical scheme. Numerical diffusivity, however, depends on the resolution of the finite grid employed. To evaluate the convergence of our results, we have performed an additional high-resolution simulation branching from the baseline setup of model M1, which we hereafter denote as model M1$_\mathrm{hr}$ (for \textit{high resolution}). Model M1$_\mathrm{hr}$ features a twofold increase in grid resolution along both the $x$- and $y$-axes, effectively quadrupling the total spatial domain sampling compared to models M1 to M6, from $1500\times2000$ points to $3000\times4000$ points in total.

A primary point of comparison between M1 and M1$_\mathrm{hr}$ is the evolution of the current density ($J_z$) distribution near the magnetic null point. Looking at the 2D spatial profile of $J_z$ in Figure \ref{fig:nullpointfieldlines_hr} and comparing with the ones in Figure \ref{fig:nullpointfieldlines}, we see a strong qualitative agreement between the M1 and M1$_\mathrm{hr}$ runs. The difference lies in that M1$_\mathrm{hr}$ resolves steeper gradients, due to its increased grid resolution and reduced numerical diffusion, resulting in higher peaks, in absolute values, of the localized current. The convergence is further demonstrated by the topology of the magnetic field lines near the null point, which evolves in an almost identical manner across both grid configurations. This implies that the underlying magnetic reconnection dynamics are robustly captured by our baseline resolution and are not artifacts of grid-scale numerical dissipation.

Similarly, an examination of the plasma velocity components, $V_x$ and $V_y$, extracted along the spine of the magnetic skeleton of the null point, shown in Figure \ref{fig:ztslits_hr}, confirms that the qualitative behaviour of the flow fields is well conserved across the two grid resolutions. For reference, the resuts for M1 are shown in Figure \ref{fig:ztslits}. While focusing along the spine of the null point, the $V_y$ component consistently maps the propagation of field-aligned propagation of slow magnetoacoustic waves, whereas $V_x$ efficiently tracks the fast magnetoacoustic modes. The similarity of our results across both resolutions confirms that our physical interpretation of the wave-plasma interaction remains intact under grid refinement.

Finally, we conducted a comparative analysis of the spectra derived from the magnetic flux conversion rates ($\dot{\Phi}_B(X)$ at $X = 7.3$ Mm and $\dot{\Phi}_B(Y)$ at $Y=5.6$ Mm) near the null point and the corresponding $V_x$ and $V_y$ velocity signals. Beyond a modest redistribution of power across the secondary high-frequency peaks in the high-resolution spectra of the magnetic fluxes, the primary periodicities identified in M1 remain almost unchanged in M1$_\mathrm{hr}$, as seen in Figure \ref{fig:parametercavity_hr}. In both models, the main and most of the secondary peaks rise above the $95\%$ confidence level. This agreement increases our level of trust towards the results of models M1 to M6.

\end{onecolumn}
\end{appendix}
\end{document}